\documentclass{ws-ijgmmp}
\usepackage{cite}
\usepackage{subfigure}
\usepackage{color}

\begin{document}

%
\catchline{}{}{}{}{}
%

\title{Transitioning late-time cosmology with the Hubble parameterization}

\author{Vinod Kumar Bhardwaj}
\address{Department of Mathematics, Institute of Applied Sciences and Humanities, GLA University, Mathura 281406, Uttar Pradesh, India\\ 
\email{dr.vinodbhardwaj@gmail.com}}

\author{Saibal Ray}
\address{Centre for Cosmology, Astrophysics and Space Science (CCASS), GLA University, Mathura 281406, Uttar Pradesh, India\\ 
\email{saibal.ray@gla.ac.in}}

\author{Kazuharu Bamba}
\address{Faculty of Symbiotic Systems Science, Fukushima University, Fukushima 960-1296, Japan\\
\email{bamba@sss.fukushima-u.ac.jp}}

\author{Akram Ali}
\address{ Department of Mathematics, College of Sciences, King Khalid University, Abha 61413, Saudi Arabia\\
\email{akali@kku.edu.sa}}

\maketitle	

\abstract{We investigate a late-time cosmological model for a homogeneous and isotropic space-time in the Rastall theory. We explore the observational constraints on the Hubble parameter by using the latest cosmological datasets such as cosmic microwave background radiation (Planck), baryon acoustic oscillations (DESI) and Type Ia Supernovae (Union 3.0). As a result, we explicitly demonstrate that the specific redshift transition occurs, namely, there happens a phase shift in the evolution of the universe from the initial deceleration era to the current accelerating phase of the cosmological scenario. Furthermore, we show that with the latest dataset of DESI-BAO clubbed with CC, CMB, and Union 3.0, the current value of the Hubble parameter is estimated as $H_0 = 66.945 \pm 1.094$, which can be compatible with the available observations.
}


\section{Introduction}
Einstein’s General Theory of Relativity (GR) stands as one of the most profound scientific achievements of the twentieth century, establishing a deep relationship between the geometry of spacetime, radiation, and matter. GR has played a pivotal role in describing both the early inflationary epoch and the present as well as future phases of the universe’s expansion. Observational results from Cosmic Microwave Background Radiation (CMBR), Wilkinson Microwave Anisotropy Probe (WMAP), Type Ia supernovae (SN Ia), Large Scale Structure (LSS), Baryon Acoustic Oscillations (BAO), Dark Energy Spectroscopic Instrument (DESI), Union 3.0 and so on collectively confirm that the current universe is undergoing accelerated expansion \cite{ref1,ref2,ref6}. Although cosmological models grounded in GR successfully reproduce many observational results \cite{ref1,ref2}, the late-time cosmic acceleration poses fundamental questions regarding the GR viability on large scales \cite{ref8}. This accelerating behavior has been attributed to an unknown component with negative pressure, known as dark energy (DE), which brought back Einstein’s cosmological constant $\Lambda$ as a possible candidate for explaining it \cite{ref8,ref9}. However, despite its success, the $\Lambda$CDM framework faces persistent theoretical challenges such as the fine-tuning problem, the cosmic coincidence issue, and the $H_0$ tension. To overcome these shortcomings, numerous dynamical DE models and modified gravity theories have been introduced, encompassing extra-dimensional and scalar field approaches \cite{ref10,ref11,ref14}.

The equivalence principle forms one of the cornerstones of GR. Nonetheless, certain theoretical and observational considerations indicate that this principle might be violated under specific conditions, which has led to increased interest in alternative gravitational theories that go beyond the framework of GR. The divergence-free energy-momentum tensor (EMT), that interacts weakly with spacetime geometry, plays a fundamental role in constructing GR and its modified versions \cite{ref17,ref18}. However, several works argue that particle creation processes could violate EMT conservation, challenging the standard condition $\nabla_{\nu} T^{\mu \nu} = 0$ \cite{ref19,ref21,ref22}. This opens the possibility of formulating gravitational theories where EMT conservation is relaxed. Consequently, modified gravity frameworks \cite{ref23,ref24,ref25} have become promising tools to address cosmological issues such as (DE) \cite{ref1,ref26} as well as dark matter (DM) problems \cite{ref27,ref28}. Any such alternative, however, must be consistent with both theoretical and observational constraints. In this direction, various modified theories of gravity have been extensively developed and investigated \cite{ref29,ref32,ref34}.

Among these alternatives, the theory proposed by Peter Rastall in 1972 \cite{ref36} has gained significant attention who basically generalized the Einstein theory. Thus the Rastall theory revisits the conservation law of the EMT, questioning its strict validity in curved spacetime even though it holds in the Minkowskian spacetime limit (or weak-field approximation scenarios). In this framework, the postulate $\nabla_{\nu} T^{\mu \nu} = 0$ is replaced by a more general condition where the divergence of the EMT is directly proportional to the gradient of the Ricci scalar, i.e., $\nabla_{\nu} T^{\mu \nu} \propto \nabla^{\mu} R$. In a precise statement, Rastall gravity \cite{ref36} allows for a non-conserved EMT characterized by a non-vanishing covariant divergence of the energy-momentum tensor, which is a generalization of general relativity. Thus, the Rastall theory introduces a coupling between geometry and matter, where curvature can influence the non-conservation of energy and momentum. 

It is argued that this coupling can also be deciphered as an effective interpretation of quantum effects in the Riemannian curved spacetime \cite{ref37}. The curvature-matter coupling mechanism allows non-minimal interactions between geometry and matter, naturally guiding to a breakdown of the conventional conservation law \cite{ref38,ref41}. Recent research has extensively explored different aspects of the Rastall theory, including its implications for DE perturbations \cite{ref42}, its relation with the scalar field of Brans–Dicke theory \cite{ref43} and its predictions regarding the inner composition of compact stars, especially neutron stars \cite{ref44}. A number of works have also analyzed different cosmic evolution stages within this theoretical framework, offering valuable observational support and theoretical interpretations \cite{ref45,ref49}. 

Interestingly, the Rastall theory is unrestricted from age and entropy related issues \cite{ref50} and, therefore, delivers interpretations for both the inflationary as well as accelerated cosmic expansion \cite{ref45,ref51}. Furthermore, Moradpour et al. \cite{ref51} argued that the non-minimally coupled interaction in between the pressureless matter part and geometrical fabric of spacetime in the Rastall theory can effectively replicate dark energy behavior, successfully explaining the observed accelerated expansion of the universe. The theory has been tested across different cosmological eras, including the inflationary, matter-dominated, and late-accelerating phases \cite{ref52}. Moreover, the cosmological model developed in Ref.~\cite{ref53} under the Rastall framework shows convergence with the $\Lambda$CDM model at late times, and the inclusion of a Rastall parameter has been suggested as a possible resolution to the initial singularity problem \cite{ref52}.

It is important to highlight that there exists an ongoing debate about whether the Rastall theory is fundamentally equivalent to GR. While both frameworks share some mathematical and conceptual similarities -- particularly in terms of the Energy–Momentum Tensor and the presence of non-minimal coupling -- there are also significant distinctions. Hence, certain studies argue in favor of equivalence \cite{ref106,ref107,Chagoya:2023mpd}, whereas others support non-equivalence \cite{ref108,ref109}. In this work, however, we treat the Rastall theory within the GR platform to examine cosmological evolution. As demonstrated later (see discussions in connection to Eqs.~(\ref{eq2}) and (\ref{eq3})), under a specific constraint on the Ricci scalar $R$, both theories can coincide. However, in Ref.~[59], it has been argued that there’s no real Rastall gravity but just a redefinition of GR’s energy momentum tensor. 

In the present study, we employ modern {\it Deep Learning} (DL) techniques -- a specialized branch of Machine Learning (ML) -- to examine the accelerating expansion phase of the present universe by means of constraining cosmological parameters. Using the \texttt{CoLFI} Python package \cite{ref54}, we estimate the parameters through Artificial Neural Networks (ANN), Mixture Density Networks (MDN), and Mixture of Gaussian Networks (MNN) \cite{ref55,ref56,ref57}. These approaches allow conditional probability distributions to be learned directly from observational data while also supporting efficient inference of posterior distributions \cite{ref54,ref55,ref56}. Additionally, hyper-ellipsoid constraint techniques are employed to improve neural network (NN) training and to achieve more precise parameter estimation \cite{ref57}.

Various experimental findings which include Ia supernovae~\cite{Perlmutter1999,Riess1998}, Cosmic Microwave Background (CMB) ~\cite{Akrami2020,Aghanim2020}, Baryon Acoustic Oscillations (BAO)~\cite{Hinshaw2009,Bennett2012}, Large Scale Structure (LSS)~\cite{Carroll2004,Coil2013} conclusively confirmed the present era of an accelerating universe (also vide Ref.~\cite{Pradhan2025} and other references therein). In the present work we therefore perform a comparative assessment of neural network–based techniques against the traditional Markov chain Monte Carlo (MCMC) method to validate the already obtained features of the late time cosmic acceleration. We show that the MNN outcomes are in strong agreement with those obtained from the MCMC method, validating the reliability and effectiveness of the DL approach. The proposed cosmological model successfully reproduces accelerated phase of the present-day universe without invoking an exotic entity, i.e., DE and overcomes the issue of the erstwhile cosmological constant. Thus, the use of observational data within the Rastall theory becomes essential for accurate modeling. Our findings demonstrate that neural network-based estimation provides a powerful substitute to the MCMC approach. The Rastall framework allows investigation of gravitational effects on cosmic expansion through ANN, MNN, and MDN algorithms. This study forms part of a broader research direction integrating cosmology with ML, illustrating how AI techniques can efficiently address complex problems related to cosmic evolution as well as expansion.

The paper is organized as follows: the solution of the Rastall field equations with logarithmic parameterization of $H(z)$ is presented in Sec.~II. Sec.~III discusses observational constraints on model parameters obtained via three different methods and datasets. Section~IV examines the model’s dynamical features, including energy density, pressure, and the deceleration parameter, as well as diagnostic tools like statefinders. The energy conditions governing the model’s physical viability are analyzed in Sec.~V. We have provided information criteria and model selection in Sec.~VI whereas a detailed discussion of various aspects of the model undertaken has been done in Sec.~VII. Sec.~VI provides a few essential concluding remarks on this work. 

\section{Model and the Field Equations in Rastall's Gravity}
Rastall's gravitational framework stands out as an interesting branch within the family of extended form of GR \cite{ref106b}. In this work, we are trying to derive feasible solutions by physically considering both the matter sector and the underlying spacetime geometry. 
Unlike GR, where the conservation equation obeys $\nabla_j T^{i j}=0$, the more generalized form in the Rastall theory is written as $\nabla_j T^{i j}=u^{i}$ \cite{ref36,ref59}. To recover the standard GR conservation law, the vector on the right-hand side must vanish. Accordingly, the vector $u^{i}$ is defined as $u^{i} = \lambda \nabla^{i}R$, where $R$ and $\lambda$ respectively denote the Ricci scalar and the coupling parameter involved in the Rastall theory. The corresponding field equations governing the Rastall theory are established as \cite{ref36,ref59}
\begin{equation}\label{eq1}
R_{ij} - \frac{1}{2}g_{ij}R + k \lambda g_{ij} R = k T_{ij},
\end{equation}
where $R_{i j}$  represents the Ricci tensor, $g_{ij}$ denotes the metric tensor, and $T_{ij}$ stands for the energy–momentum tensor (EMT). The parameter $k$ is the gravitational coupling constant, whose value is fixed by ensuring consistency with the Newtonian limit. The standard Einstein field equations are recovered when $\lambda = 0$ and $k= 8\pi$, under the condition of covariant conservation of the energy–momentum tensor, $T^{i j}_{; i} = 0$. 

Hence, the trace involved in Eq.~(\ref{eq1}) can easily be presented in the following form \cite{ref59}
\begin{equation}\label{eq2}
		(4 k \lambda - 1) R = k T,
\end{equation}
from where one can notice that the energy–momentum tensor becomes traceless (i.e., $T = 0$) when either $R = 0$ or $k\lambda = \frac{1}{4}$. Hence, it is important to highlight that the Rastall theory exhibits behavior analogous to GR in the case of a vanishing Ricci scalar ($R = 0$), while the EMT trace disappears when $k\lambda = \frac{1}{4}$. When $k\lambda \neq \frac{1}{4}$, the gravitational field equations incorporating the cosmological constant $\Lambda$ can be expressed as follows
\begin{equation}\label{eq3}
G_{ij} +\Lambda g_{ij}+ k \lambda g_{ij} R = k T_{i j},
\end{equation}
where $G_{ij}$ represents a typical tensor after the name of Einstein.

The non-flat metric, i.e., the Friedmann-Lema\^{i}tre-Robertson-Walker(FLRW) line element is adopted here to describe a universe in its usual format which can be given by
\begin{equation}\label{eq4}
ds^{2} = -dt^2+a^{2}(t) \bigg(\frac{dr^2}{1-K r^2}+r^2 d\Omega^2\bigg),
\end{equation}
where $d\Omega^2 = d\theta^2+\sin^{2}\theta d\phi^2$. It is to note that the cosmic expansion is governed by the scale factor $a(t)$, which depends on time, while the spatial curvature index $K$ specifies the geometry of the universe, where $K = -1$, $0$, $+1$ represent open, flat, and closed spatial configurations, respectively.

The non-vanishing components are: \\
\(R_{00} = -3\frac{\ddot{a}}{a} = -3(\dot{H} + H^2)\); \, 
\(R_{ij} = g_{ij} \left( \frac{\ddot{a}}{a} + 2\frac{\dot{a}^2}{a^2} + \frac{2K}{a^2} \right) = g_{ij} (\dot{H} + 3H^2 + \frac{2K}{a^2})\);\\
\(R = 6 \left( \frac{\ddot{a}}{a} + \frac{\dot{a}^2}{a^2} + \frac{K}{a^2} \right) = 6(\dot{H} + 2H^2 + \frac{K}{a^2})\).

For perfect fluid: $u^i = (1, 0, 0, 0)$; \, $T_{00} = \rho$,
\, $T_{ij} = p g_{ij}$. 

For $i=0$ and $j=0$, from Eq. (\ref{eq1}) we get 
\begin{equation}\label{eq2.5}
R_{00} - \frac{1}{2}g_{00}R + k \lambda g_{00} R = k T_{00}
\end{equation}
Since  $g_{00} = -1$, then 
\begin{equation}\label{eq2.6}
-3(\dot{H} + H^2) + \frac{1}{2}R - k \lambda R = k \rho
\end{equation}

Using the relation $R = 6 \left( \frac{\ddot{a}}{a} + \frac{\dot{a}^2}{a^2} + \frac{K}{a^2} \right) = 6(\dot{H} + 2H^2 + \frac{K}{a^2})$, 
\begin{small}
\begin{equation}\label{eq2.7}
-3(\dot{H} + H^2) + 3(\dot{H} + 2H^2 + \frac{K}{a^2}) - 6k \lambda (\dot{H} + 2H^2 + \frac{K}{a^2}) = k \rho
\end{equation}
\end{small}

Again, we get the following expression on dividing Eq.~(\ref{eq1}) by  $g_{ij}$ to work with scalars:
\begin{small}
\begin{equation}\label{eq2.8}
(\dot{H} + 3H^2 + \frac{2K}{a^2}) - 3(\dot{H} + 2H^2 + \frac{K}{a^2}) + 6k \lambda (\dot{H} + 2H^2 + \frac{K}{a^2}) = k p.
\end{equation} 
\end{small}

On simplifying Eqs. (\ref{eq2.7}) and (\ref{eq2.8}), the field equations in the Rastall theory can be written as \cite{ref36,ref59}
\begin{equation}\label{eq5}
(3 - 12 k \lambda) H^2 - 6 k \lambda \dot{H} + (3 - 6k\lambda) \frac{K}{a^2} = k \rho,
\end{equation}

\begin{equation}\label{eq6}
 -(3 - 12k\lambda) H^2 - (2 - 6k\lambda) \dot{H} - (1 - 6k\lambda) \frac{K}{a^2} = k p,
\end{equation}
where $p$ denotes the cosmic pressure and $\rho$ represents the energy density while the symbol over-dot denotes differentiation under the cosmic time $t$ whereas $a$ is the scale factor which is attached to the Hubble parameter via the relation $H=\dot{a}/a$. 

For the considered model, the cosmic fluid pressure and matter-energy density are associated via the barotropic equation of state $p= \omega \rho$, with symbol $\omega$ as the EOS parameter. Using the Bianchi identity, expressed as $G^{; j}_{ij} = 0$, one can obtain the energy-momentum conservation (continuity) equation \cite{ref59}
\begin{equation}  \label{eq7}
(1 - 3 \lambda k) \dot{\rho} - 3 \lambda k \dot{p}+ (3- 12 \lambda k) (p+\rho) H = 0.
\end{equation}

Most of the cosmological observational data are expressed as function of the redshift $z$ instead of $t$. Typically the cosmic scale factor $a(t)$ is related to $z$ via the $a_{0}/a = 1+z$ relationship with $a_{0}$ as the presently considered value of the scale factor (which we have assumed as unity in the present study). By using this relation, the time derivative of the Hubble parameter can be presented in the following form: $\dot{H}(t)=-(1+z) H(z) H'(z)$, as such $H'(z)= \frac{dH}{dz}$. Consequently, the field equations given in Eqs.~(\ref{eq5}) and~(\ref{eq6}) can be rewritten in terms of the redshift parameter $z$ as follows
\begin{eqnarray}\label{eq8}
& (3 - 12 k \lambda) H^2 + 6 k \lambda (1+z) H H'(z) \nonumber\\
&+ K  (3 - 6 k \lambda k) (1+z)^2 = k \rho,
\end{eqnarray}
\begin{eqnarray}\label{eq9}
& (3 - 12 k \lambda) H^2 + (1+z) H (6 k\lambda - 2) H'(z)\nonumber\\
&- K (6 k \lambda - 1 )  (1+z)^2  = - k p.
\end{eqnarray}

To explain DE related cosmological phenomena, various theoretical models have been proposed in literature, however none of them could explain it in a significant and efficient way. To understand the physical dynamics of the universe, therefore we need parameterization. However, a model-independent technique is the well-known alternative for verifying the dynamics of DE models \cite{ref60}. This strategy aims to follow the parameterization of the any one of the cosmic parameters, i.e., the Hubble parameter, Deceleration Parameter (DP), and EOS parameter, as required for any auxiliary equations. Considering the parameterization of $H(z)$, the state-finder diagnostic has been examined in Ref. \cite{ref61}. In addition, several parameterizations have been attempted by different authors for exploration of the dynamic attributes of our universe which include CPL (Chevallier-Polarski-Linder), BA (Barboza-Alcaniz), LC (Low Correlation) \cite{ref71}, and the DP  \cite{ref72}. 

In the literature, therefore the above mentioned parameterizations have been extensively considered for issues with cosmological researches. Essentially the main motivation we would like to adhere here is to allow the considered parameterization for deviations from the standard $\Lambda$CDM model at both low and high redshifts. Under such motivation, we consider here parameterization of the Hubble parameter $H(z)$ in the logarithm form as \cite{ref74}
\begin{equation} \label{eq10}
H^{2}(z) = H^{2}_{0} \bigg[(z+1)^3 A + B + \eta  \log (1+z)\bigg],
\end{equation}
where $H_{0}$ is the current value of Hubble constant. The parameterization Eq.~(\ref{eq10}) can be reduced into the $\Lambda$CDM cosmology if $A = \Omega_\mathrm{m0}$, $B = (1-\Omega_\mathrm{m0})$, and $\eta = 0$. However, the addition of the term $\eta \log(1+z)$ provides a kind of logarithmic modification that is introduced to enhance the model’s adaptability in describing more intricate features of the cosmic expansion history. This formulation is motivated by the objective of examining possible departures from the $\Lambda$CDM framework, which relies on a constant cosmological term $\Lambda$ along with a fixed matter density parameter. By incorporating the logarithmic component, an extra adjustable parameter is introduced, allowing the model to capture evolution patterns in the expansion rate that may not be fully represented by simpler parameterizations. This kind of generality is useful to test the model against observational constraints and evaluating its potential to more accurately reflect the actual expansion dynamics of the universe \cite{ref75,ref76}.

In the above Eq.~(\ref{eq10}), the normalizing feature $H_{0}$ confirms that the free parameters $A$ and $\eta$ are dimensionless and of order unity. This parameterization enables us to analyze the dynamics of a cosmic model beyond the conventional $\Lambda$CDM model across both high and low redshift ranges. Later on one can observe that by using the same parametrization as of Ref.~\cite{ref74}, a better model is possible to provide based on data analytic approaches which does altogether represent a valid scenario of late time accelerated universe. 

From Eq.~(\ref{eq10}), we get the relation $A+B = 1$ at $z=0$. Within the $\Lambda$CDM cosmological framework, the Hubble parameter is expressed as follows
\begin{equation} \label{eq18}
H(z) = H_{0} \sqrt{\Omega_{\mathrm{m}0} (z+1)^3 +\Omega_{\Lambda}}.
\end{equation}
where $\Omega_{m0}$ and $\Omega_{\Lambda}$ are the current values of density parameters of matter and dark energy. As we are treating Rastall gravity as an extension of GR, so here $\Omega_{\Lambda}$ taking the role of Einstein's erstwhile cosmological constant $\Lambda$ based cosmic density which dynamically acts as an agent for the expansion of the universe. 

For the considered model, the DP can be written as
\begin{equation}\label{eq19}
q =  - H^{-2} \dot{H} - 1. 
\end{equation}

\section{Methodologies and datasets}

In this section, we perform data analysis under statistical as well as ML based discussion. 

\subsection{Statistical analysis}

\textbf{A-1: Statistical analysis by using the primary observational datasets}

Bayesian inference has emerged as a cornerstone methodology in contemporary cosmological research for parameter estimation and model comparison. In this investigation, we aim to determine constraints on the free parameters of our cosmological framework by analyzing various observational datasets through a $\chi^2$-minimization approach. To derive estimates for the model parameters $A$, $B$, $\eta$, and $H_{0}$, we utilize the following observational datasets:

\textbf{(i) Cosmic Chronometers (CC):}  
The cosmic chronometer technique offers direct observational measurements of the Hubble parameter $H(z)$ derived from the differential age evolution analysis of galaxies. Our analysis incorporates a compilation comprising 31 observational data points distributed across the redshift range [0.07, 1.965] (vide refs. \cite{ref77,Moresco2015}).

The $\chi^{2}$ function to comprise the covariance matrix is read as:
\[
\chi^2_{\mathrm{CC}} = (\mathbf{H}_{obs} - \mathbf{H}_{model})^T \cdot \mathbf{C}^{-1} \cdot (\mathbf{H}_{obs} - \mathbf{H}_{model}),
\]
where $\mathbf{H}_{obs}$ and $\mathbf{H}_{model}$ denote the observed and theoretical $H(z)$ values, respectively, and $\mathbf{C}$ represents the covariance matrix. The covariance matrix for $H(z)$ measurements is given as $C = C^{stat}+C^{sys}$, where $C^{stat}$ is known as statistical covariance which is usually a diagonal matrix representing the observational uncertainties of each $H(z)$ measurement and $C^{sys}$ is called systematic Covariance that consists the systematic correlated error in CC measurements by shared physics and calibration. 

\textbf{(ii) Pantheon sample:}  
Here, we employ the most latest sample of SN Ia within the redshift interval [0.001, 2.26] \cite{ref80}. It is to be noted that this dataset integrates observations from around 18 distinct surveys which include all light curves normalized by  using the SALT2 fitting procedure and thus to maintain consistent calibration throughout the dataset \cite{ref81}.

For Ia Supernovae (SNe Ia) sample, the $\chi^2$ statistic is defined as
\begin{equation}
\chi^2 = \Delta {\mu}^T \mathbf{C}^{-1} \Delta {\mu},
\end{equation}
where $\Delta {\mu} = {\mu}_{obs} - {\mu}_{th}(\theta)$ is the residual vector, $\mathbf{C}$ is the covariance matrix, and  $\theta$ represents cosmological parameters. The covariance matrix C for Pantheon and Pantheon+  samples is generally sum of two primary components: (i) $C_{stat}$ which is a diagonal-dominant matrix containing the measurement uncertainties for each SN. This includes the uncertainty from the light-curve fit (SALT2 model), the ``intrinsic scatter" of SNe Ia, and lensing/peculiar velocity uncertainties, (ii) $C_{sys}$ which is a non-diagonal matrix that captures how errors are correlated across the SNe Ia sample.  

\textbf{(iii) Baryon Acoustic Oscillation (BAO) Data:}  
Our analysis incorporates six BAO measurements derived from various galaxy surveys, such as SDSS DR7, 6dF, and WiggleZ \cite{ref82,ref83,ref84} (see Table \ref{tab:t1}). Here the angular diameter distance can be provided in the following form 
\begin{equation}
d_\mathrm{A} = \frac{D_\mathrm{L}}{(1+z)^{2}},
\end{equation}
where $D_\mathrm{L}$ denotes the luminosity distance \cite{ref85}. 

Again, the scale of dilation is formulated as  

\begin{equation}
D_\mathrm{V}(z) = \left[ D_\mathrm{L}^{2}(z) (1+z)^{-2} \frac{cz}{H(z)} \right]^{1/3}.
\end{equation}

For constraining the model parameters, we define the chi-square estimator for the BAO observations as \cite{ref85,ref86,ref87,ref88}
\begin{equation}
\chi^{2}_{BAO} = X^{T} C^{-1} X,
\end{equation}
where $C$ represents the covariance matrix and the vector $X$ is given by
\begin{align*} \label{X}
X &= \begin{pmatrix}
\frac{d_A(z_{*})}{D_V(0.106)} - 30.95 \\
\frac{d_A(z_{*})}{D_V(0.20)} - 17.55 \\
\frac{d_A(z_{*})}{D_V(0.35)} - 10.11 \\
\frac{d_A(z_{*})}{D_V(0.44)} - 8.44 \\
\frac{d_A(z_{*})}{D_V(0.60)} - 6.69 \\
\frac{d_A(z_{*})}{D_V(0.73)} - 5.45
\end{pmatrix}.
\end{align*}

and $C^{-1}$ is given by \cite{ref85}
\begin{table}[ht]
\small
\begin{align}
C^{-1} &= \begin{pmatrix}
		0.48435 & -0.101383 & -0.164945 & -0.0305703 &-0.097874 & -0.106738\\           
		-0.101383 & 3.2882 &-2.45497 &-.0787898 &-0.252254 &-0.2751\\
		-0.164945 &-2.454987 &9.55916 &-0.128187 &-0.410404 &-0.447574\\
		-0.0305703 &-0.0787898 &-0.128187 &2.78728&-2.75632 &1.16437\\
		-0.097874 &-0.252254 &-0.410404 &-2.75632 &14.9245 &-7.32441\\
		-0.106738 &-0.2751 &-0.447574 &1.16437 &-7.32441 &14.5022
		\end{pmatrix}.
\end{align}
\label{15}
\end{table}
One can evaluate the parametric values of the model, such as $A$, $B$, $\eta$, and $H_{0}$, through the observational evidences and statistical techniques. Here, $E_\mathrm{obs}$ denotes the measured value from observational data, whereas $E_\mathrm{th}$ signifies the theoretically predicted values within our cosmological framework. Through statistical comparison of $E_\mathrm{obs}$ and $E_\mathrm{th}$, we can determine the optimal parameter values. For achieving the most precisely obtained values, we execute the $\chi^2$ estimator as follows
\begin{equation} \label{chi1}
\chi^{2} = \sum_\mathrm{i=1}^{N}\left[\frac{E_\mathrm{th}(z_{i})- E_\mathrm{obs}(z_{i})}{\sigma_\mathrm{i}}\right]^{2},
\end{equation}
where $E_\mathrm{th}(z_{i})$ presents the theoretical parametric values, $E_\mathrm{obs}(z_{i})$ denotes the measured values, $\sigma_{i}$ indicates the associated uncertainty (i.e., the standard error), where $N$ stands for the total points of data employed in the sample.
	
The combined $\chi^{2}$ estimator can be given by
\begin{equation}
\label{chi2}
\chi^{2}_\mathrm{combined} = \chi^{2}_\mathrm{CC} + \chi^2_\mathrm{Pantheon} + \chi^{2}_\mathrm{BAO}.
\end{equation}

\begin{table}[ht]
	   \centering
       \small
 \setlength{\tabcolsep}{5pt} 
    \renewcommand{\arraystretch}{1.0} 
	    \caption{Datasets of $z_\mathrm{BAO}$ and $\varphi(z)$ to evaluate $\Upsilon (z)$. Here, we exploit the relationship $\varphi(z)= d_A(z_{*})/D_V(z_\mathrm{BAO})$ with $z_{*}\approx 1091$~\cite{ref85}).}
	 	\label{tab:t1}
	     \begin{tabular}{ccccccc}
    \hline \hline 
 
$z_\mathrm{BAO}$ & $0.106$ & $0.2$	 & $0.35$  & $0.44$ & $0.6$ &$ 0.73$  \\
		
$\varphi(z)$ &  \small$30.95 \pm 1.46$ & \small$17.55 \pm 0.60$  & \small$10.11 \pm 0.37$ & \small $8.44 \pm 0.67$ &\small $6.69 \pm 0.33$ & $ 5.45 \pm 0.31$\\
\hline
 \end{tabular}
\end{table}

In the context of the $\chi^{2}_\mathrm{tot}$  statistics, a few technical points are as follows that one can make it minimized in order to get the parametric value which could be best fitted to  the  datasets of the joint sample of CC, Pantheon, and BAO. Therefore, (i) under the approach of the maximum likelihood, $\mathcal{L}_\mathrm{tot} = exp \left(-\chi^2_\mathrm{tot}/2 \right)$, known as the total likelihood function and can be expressed as: $\mathcal{L}_\mathrm{tot}=  \mathcal{L}_\mathrm{BAO}*\mathcal{L}_\mathrm{Pantheon}* \mathcal{L}_\mathrm{CC}$; (ii) either the likelihood function $\mathcal{L}_\mathrm{tot}(x*)$ can be considered as maximized or $\chi^2_\mathrm{tot} (x^{*})=-2 \ln \mathcal{L}_\mathrm{tot} (x^{*})$ can be assumed as minimized so that the most plausible values of parameters can be obtained; (iii) within the setting of the constraints (such as $x^{*}$, $1 \sigma$ and $2 \sigma$), we observe that parameters are fitted and also restricted for $\chi^2_\mathrm{tot} (x)=\chi^2_\mathrm{tot} (x^{*})+2.3$ and $\chi^2_\mathrm{tot} (x)=\chi^2_\mathrm{tot} (x^{*})+6.17$; (iv) one can get the best-fit parametric values by minimizing the $\chi^2$ statistics (see Table \ref{tab:t3} along with the illustrated Fig. 1).\\

\begin{table}[ht]
\centering
\small
\setlength{\tabcolsep}{5pt} 
\renewcommand{\arraystretch}{1.0} 
\caption{The present value of the Hubble parameter. Here, $M_1$ and  $M_2$ are the representatives of CC and BAO methods, respectively.}
\label{tab:t2}
\begin{tabular}{cccc}
\toprule 
 
$z$ &  $H_{0}$ (km~$\mathrm{s^{-1}}$~$\mathrm{Mpc^{-1}}$) &  Method & Reference \\
\hline    
0.07 & $69 \pm 19.6$ & $M_1$ & \cite{Zhang2014}\\
0.09 & $69 \pm 12$ & $M_1$ &  \cite{Simon2005}\\
0.12 & $68.6 \pm 26.2$ & $M_1$ & \cite{Zhang2014}\\
0.17 & $83 \pm 8$ & $M_1$ & \cite{Simon2005}\\
0.1791 & $75 \pm 4$ & $M_1$ & \cite{Moresco2012}\\
0.1993 & $75 \pm 5$ & $M_1$ & \cite{Moresco2012}\\
0.2 & $72.9 \pm 29.6$ & $M_1$ & \cite{Zhang2014}\\
0.27 & $77 \pm 14$ & $M_1$ & \cite{Simon2005}\\
0.28 & $88.8 \pm 36.6$ & $M_1$ & \cite{Zhang2014}\\
0.352 & $83 \pm 14$ & $M_1$ & \cite{Moresco2012}\\
0.38 & $81.9 \pm 1.9$ & $M_2$ & \cite{Alam2017}\\
0.3802 & $83 \pm 13.5$ & $M_{1}$ & \cite{Moresco2016}\\
0.4 & $95 \pm 17$ & $M_{1}$ & \cite{Simon2005}\\
0.4004 & $77 \pm 10.2$ & $M_{1}$ & \cite{Moresco2016}\\
0.4247 & $87.1 \pm 11.2$ & $M_{1}$ & \cite{Moresco2016}\\
0.44497 & $92.8 \pm 12.9$ & $M_{1}$ & \cite{Moresco2016}\\
0.4783 & $80.9 \pm 9$ & $M_{1}$ & \cite{Moresco2016}\\
0.48 & $97 \pm 62$ & $M_{1}$ & \cite{Ratsimbazafy2017}\\
0.593 & $104 \pm 13$ & $M_{1}$ & \cite{Moresco2016}\\
0.68 & $92 \pm 8$ & $M_{1}$ & \cite{Moresco2016}\\
0.781 & $105 \pm 12$ & $M_{1}$ & \cite{Moresco2016}\\
0.875 & $125 \pm 17$ & $M_{1}$ & \cite{Moresco2016}\\
0.88 & $90 \pm 40$ & $M_{1}$ & \cite{Ratsimbazafy2017}\\
0.9 & $117 \pm 23$ & $M_{1}$ & \cite{Simon2005}\\
1.037 & $154 \pm 20$ & $M_{1}$ & \cite{Moresco2016}\\
1.3 & $168 \pm 17$ & $M_{1}$ & \cite{Simon2005}\\
1.363 & $160 \pm 33.6$ & $M_{1}$ & \cite{Moresco2015}\\
1.43 & $177 \pm 18$ & $M_{1}$ & \cite{Simon2005}\\
1.53 & $140 \pm 14$ & $M_{1}$ & \cite{Simon2005}\\
1.75 & $202 \pm 40$ & $M_{1}$ & \cite{Simon2005}\\
1.965 & $186.5 \pm 50.4$ & $M_{1}$ & \cite{Moresco2015}\\
\hline
\end{tabular}
\end{table}

\vspace{2.50cm}

	\begin{figure}
		\centering
		\includegraphics[width=0.5\textwidth]{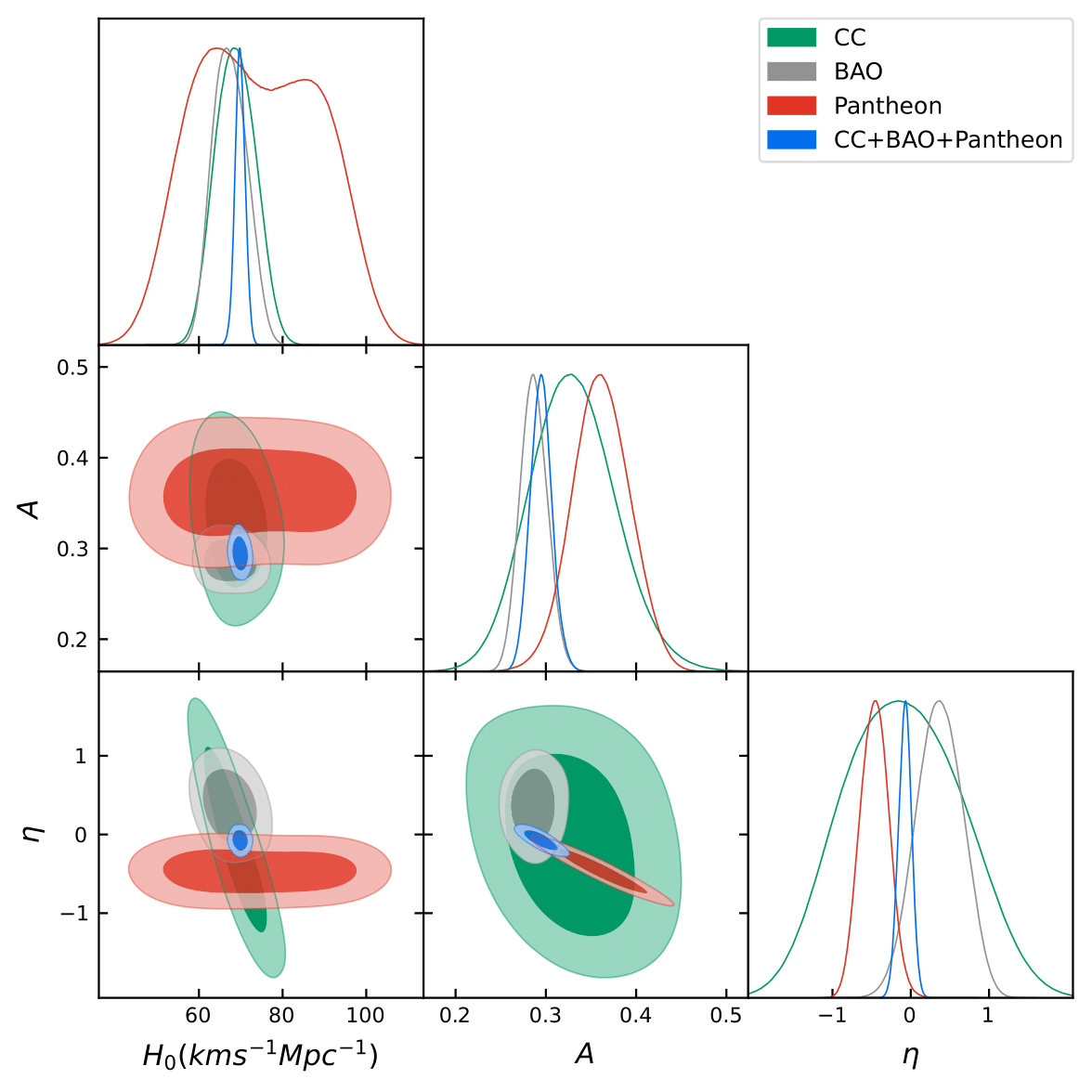}
		\caption{Contour based on the model parameters with $1\sigma$ and $2\sigma$ confidence levels for CC, Pantheon, BAO, and joint datasets.}
	\end{figure}

{{\bf {A-2 : } {Statistical analysis with the latest datasets (CMB, BAO-DESI and Union 3.0)}}
    
Here, we conduct analysis with the latest datasets of CMB, BAO-DESI, and Union 3.0 in addition with CC dataset \cite{Rubin2025SNe,Tutusaus2019SNeIa,Benevento2020Hubble,Chen2019planck,Aghanim2018cmb,Adame2025desi,Lodha2025desi,Karim2025desi} which can be seen in Fig. 2 (also inquisitive readers may be interested in Table \ref{tab:t4} for a few collected numerical values on the Hubble parameter derived from CC and BAO methodology).\\

\textbf{(i) Union 3.0}: In the present analysis, the Union 3.0 latest dataset derived from the recent SN Ia. This is a largest dataset compiled from 24 different samples of Type Ia supernovae by an unified Bayesian process and consists of 2087 SNIa within the redshift range $0.01 < z <2.26 $ \cite{Rubin2025SNe}. It is to clarify here that in the present analysis, we employ available binned dataset which is based on the distance moduli for this sample.  It to be noted that the provided dataset is based on the observation of apparent magnitude ($m$). However, the distance modulus can then be given by 
\begin{equation}
\mu (z) = m_\mathrm{B} (z) - M_\mathrm{B} =5 \log_{10} \bigg(\frac{d_\mathrm{L} (z)}{\mathrm{Mpc}}\bigg)+25,
\end{equation}
with $d_\mathrm{L}$ as the luminosity distance which can be provided as
\begin{equation}
d_\mathrm{L} (z) = c (1+z) \int_{0}^{z} \frac{dz'}{H(z')}.
\end{equation}
  
However, there are argument that the luminosity (or absolute magnitude) of SN Ia may not evolve with the cosmic redshift \cite{Tutusaus2019SNeIa} and therefore, the SN Ia sample is observed to distributed over a mean absolute magnitude for which MB = -19.22 \cite{Benevento2020Hubble}.

In a standard MCMC, the Union 3.0 dataset provides distance moduli $\mu_{\text{obs}}$ and a systematic covariance matrix $\mathbf{C}$. Then, the $\chi^2$ function is given as:
\begin{equation}
\chi^2(\theta) = \sum_{i,j} \left[ \mu_{obs, i} - \mu_{th}(z_i, \theta) \right] (\mathbf{C}^{-1})_{ij} \left[ \mu_{obs, j} - \mu_{th}(z_j, \theta) \right].
\end{equation}

The primary covariance matrix for Union 3.0 consists of (i) Statistical uncertainties: Individual light-curve measurement errors, (ii) Systematic uncertainties: Photometric calibration (zeropoints, filter passbands), host-galaxy mass corrections, and dust extinction modeling, and (iii) Correlations: Common calibration sources across different surveys and peculiar velocity correlations.

Unlike other compilations that relied on a simple $\chi^2$ approach with a pre-computed systematic covariance matrix, Union 3.0 is built using the UNITY 1.5 (Unified Bayesian Inference for Type-Ia Supernovae) framework. This means the full covariance is often handled internally through hierarchical modeling. \\

{\textbf{(ii) CMB}}: In the present investigation, we employ the CMB distance priors as available in the following work \cite{Chen2019planck} which have been obtained from the Planck 2018 temperature and polarization data (i.e., TT, TE, and EE + lowE) \cite{Aghanim2018cmb}.  Essentially, these priors encode a few crucial geometric information from the entire spectrum of CMB which is accompanied with the shift parameter (i.e., $R_\mathrm{shift}$), the acoustic scale (i.e., $l_\mathrm{A}$), and the baryon density (i.e., $\Omega_\mathrm{b}h^2$). All of them permit us to incorporate CMB constraints and the fact is that one can do this operation even without estimating the entire set of perturbative procedures.
 
Therefore, one can incorporate all the aforesaid parameters within the CMB distance prior dataset, i.e., $R_\mathrm{shift}$, $l_\mathrm{A}$, and $\Omega_\mathrm{b}h^2$ which can be evaluated as
\begin{equation}
	l_\mathrm{A} =(1+z_{*}) \frac{\pi D_\mathrm{A} (z_{*})}{r_\mathrm{s}(z_{*})},~R_\mathrm{shift} = \frac{(1+z_{*}) D_\mathrm{A} (z_{*})\sqrt{\Omega_m H^2_{0}}}{c},
\end{equation}
where $z_{*}$ and $r_\mathrm{s}(z_{*})$ are, respectively, the redshift at the decoupled epoch of photon and the co-moving sound horizon at the defined epoch.

The chi-square estimator for the CMB dataset is described in the following form \cite{Hu1996pert}
\begin{equation}
	\chi^2_{CMB} = X_{CMB}^{T} C_{CMB}^{-1} X_{CMB},
\end{equation}
where 
\begin{align*}
	X_{CMB} &= \begin{pmatrix}
		l_{a}-302.40\\           
		R-1.7264\\
		z_{*}-1090.88
	\end{pmatrix},
\end{align*}
and 
\begin{align*}
	C_{CMB}^{-1} &= \begin{pmatrix}
		3.182 & 18.253 & -1.429 \\           
		18.253 & 11887.879 &-193.808 \\
		-1.429 &-193.808 & 4.556 
	\end{pmatrix}.
\end{align*}}

\textbf{(iii) BAO-DESI}: For the analysis, we use the BAO-DESI data consists of 12 BAO samples obtained from observations on various tracers of the Dark Energy Spectroscopic Instrument (DESI) \cite{Adame2025desi,Lodha2025desi,Karim2025desi}. The observable measurements in this dataset are $d_\mathrm{M}(z)/r_\mathrm{d}$, $d_\mathrm{H}(z)/r_\mathrm{d}$, and $d_\mathrm{V}(z)/r_\mathrm{d}$, where $d_\mathrm{M}(z)= d_\mathrm{L}(z)/(1+z)$, $d_\mathrm{H}(z)= c/H(z)$, $d_\mathrm{V}(z)=\left[z d_\mathrm{H} (z) d^2_\mathrm{M}(z)\right]^{1/3}$, and $r_\mathrm{d}$ are, respectively, the co-moving distance (transverse type), the Hubble distance, the spherically angular diameter (averaged distance) and the sound horizon scale at the baryon drag epoch as a cosmological standard ruler.  

The $\chi^2$ function for DESI dataset are defined as 
\begin{equation}
\chi^2_{\text{DESI}} = \sum_{i,j} (\mathbf{y}_{\text{data}} - \mathbf{y}_{\text{model}})_i^T \mathbf{C}_{ij}^{-1} (\mathbf{y}_{\text{data}} - \mathbf{y}_{\text{model}})_j
\end{equation}

Here, the covariance matrix  $\mathbf{C}$ involves the statistical correlation between the 7 primary redshift bins spanning $0.1 < z < 4.2$, Cross-correlations between tracers that occupy the same volume (the LRG and ELG samples at $z \approx 0.8$), and systematic uncertainties were extensively tested, they were found to be significantly smaller than statistical errors ($\sigma_{\text{sys}} < \sigma_{\text{stat}}/3$) and are often neglected in the final covariance matrices.

\begin{figure}
	\centering
	\includegraphics[width=0.5\textwidth]{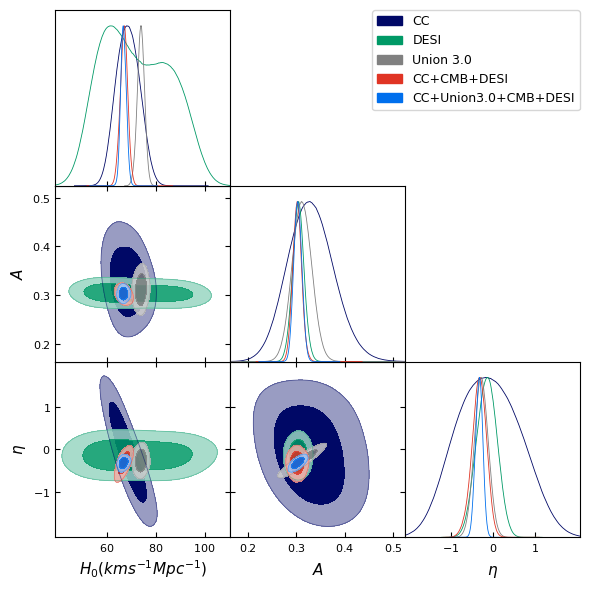}
	\caption{Contour based on the model parameters with $1\sigma$ and $2\sigma$ confidence levels for CC, Union 3.0, CMB, DESI, and joint datasets.}
\end{figure}

\subsection{Machine Learning technique}
    \begin{figure}
		\centering
(a) \includegraphics[width=0.5\textwidth]{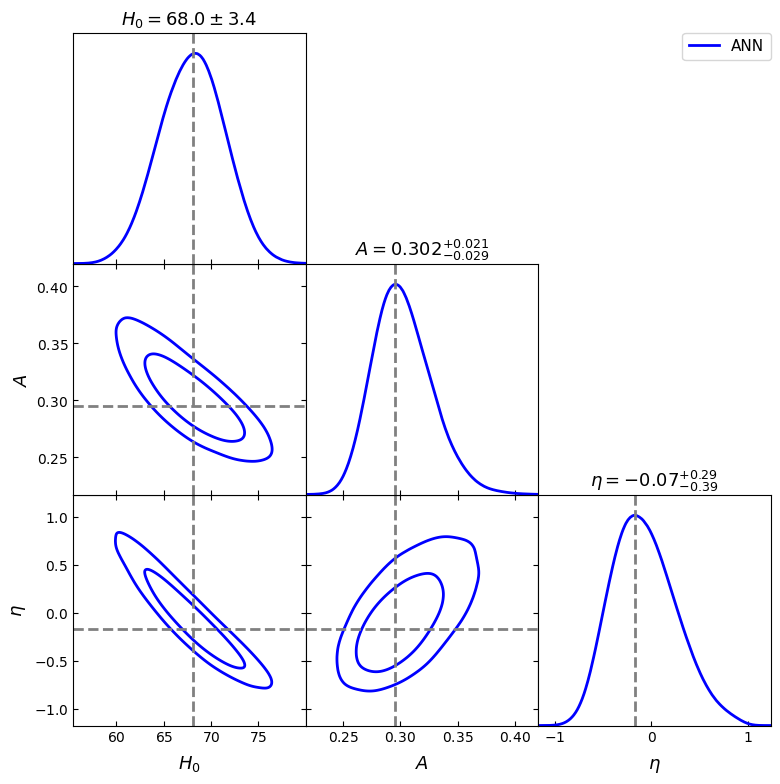}\\
(b) \includegraphics[width=0.5\textwidth]{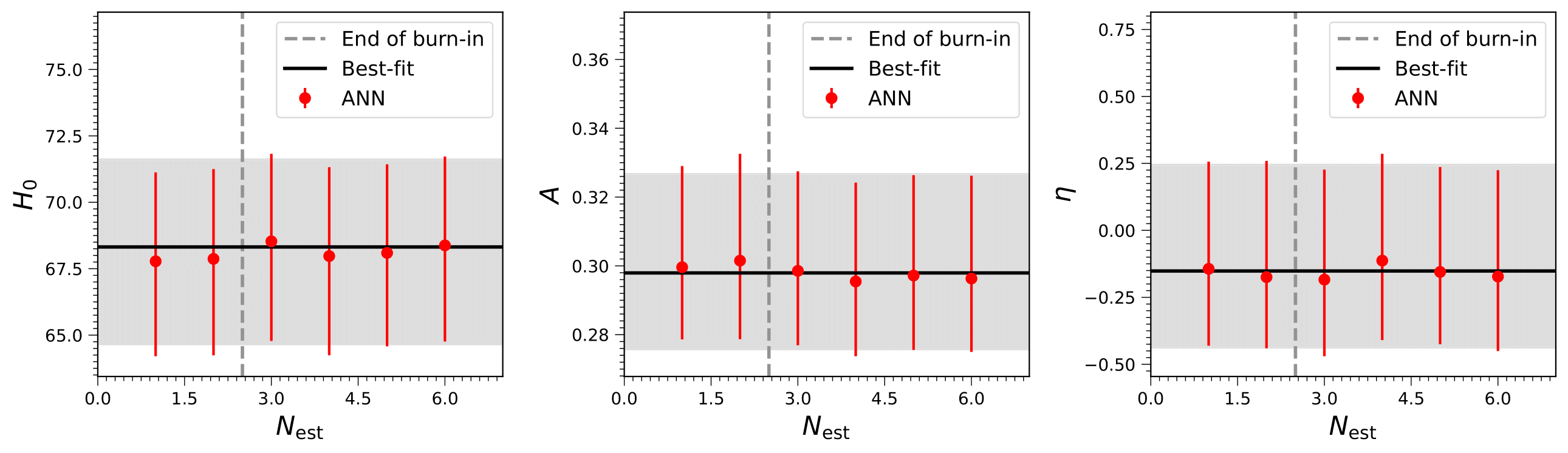}\\
(c) \includegraphics[width=0.5\textwidth]{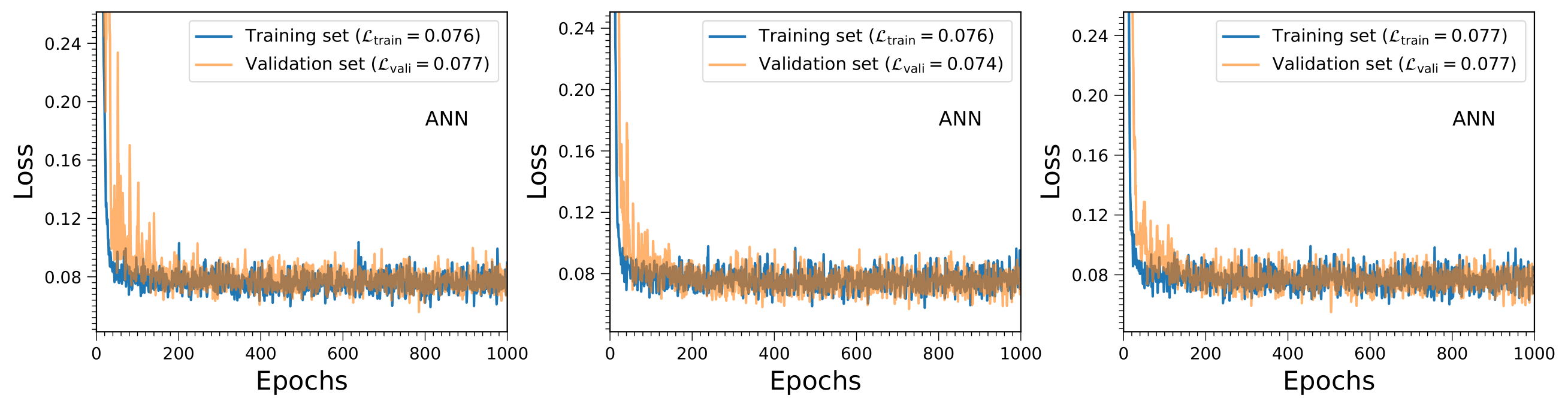}
\caption{Observational investigation of the Rastall theory based cosmology under ANN model: (a) contours drawn from $H(z)$ data for $A$, $\eta$, and $H_{0}$ $1\sigma$ and $2\sigma$, (b) relationship in between steps as well as best-fit values within $1 \sigma$ error of model parameters (where the solid black line and Grey-shaded parts display the best-fit values, while the red circle with error bars), and (c) graphical presentation of losses regarding the training and validation sets. Here, the training set consists of 2000 samples whereas the validating set contains 500 samples.}
\end{figure}

Nowadays, it is noted that due to the advancement in computing, neural network is gradually becoming a very effective and relevant tool leading to the emergence of a novel data analytic arena known as Deep Learning (DL), which acts as a subset of ML (already mentioned earlier) that uses multi-layered artificial neural networks. In the present investigation, we are thinking to apply the deep neural network in its fundamental aspect which is normally known as the multilayer perceptron (MLP). However, we have treated the performances with only the primary datasets for CMB, BAO, DESI and Union 3.0. We have checked that, apparently, the behavior and looking are too similar to distinguish each figures on ANN, MDN, and MNN for the primary as well as the latest datasets. Therefore, we consider all the figures based on primary datasets (as being important figures).  

We have used 1$e^{-2}$ as the learning rate for the training of the neural network. We utilize 2000 epochs for the analysis with a batch size consisting of 100 data points in each epoch. Here, the epoch refers to the number of times the algorithm sees the data. Further, we adopt the train and test split method for the analysis, where we employ $75\%$ data points for training and $25\%$ for validating the proposed algorithm. Furthermore, we have used 3 chains with three hidden layers for the analysis and `softplus' activation function has been utilized for each neuron in the neural network.\\

\textbf{B-1: Analysis with the primary datasets}\\

\begin{figure}
\centering
(a) \includegraphics[width=0.5\textwidth]{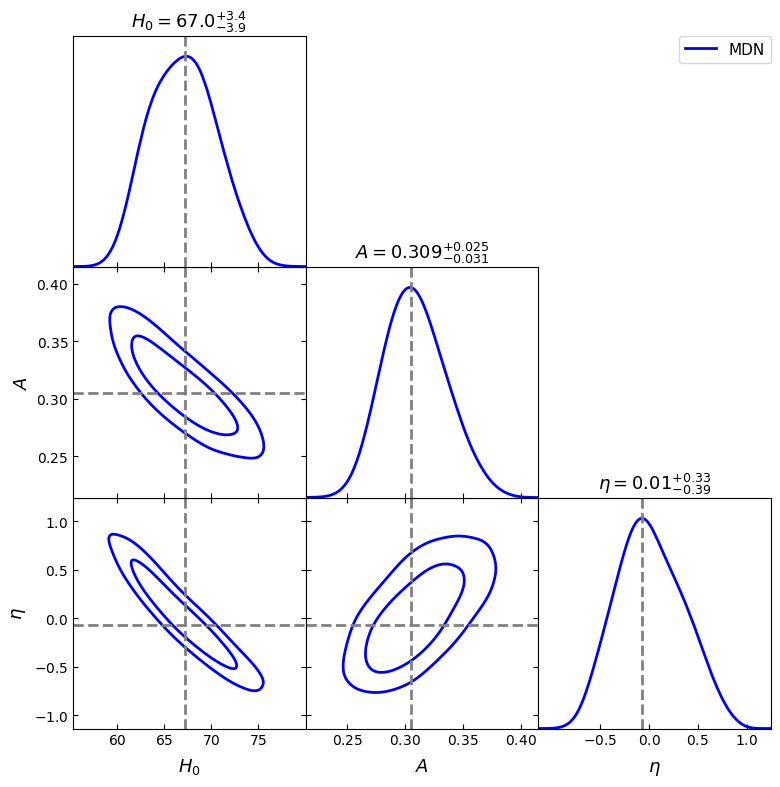}\\
(b) \includegraphics[width=0.5\textwidth]{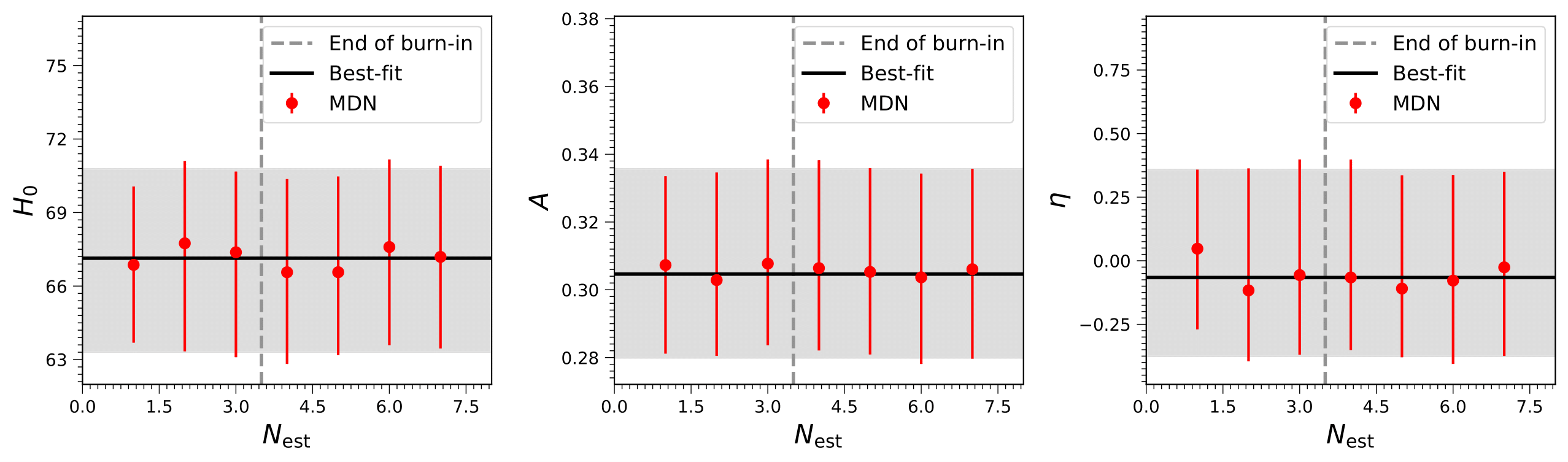}\\
(c)	\includegraphics[width=0.5\textwidth]{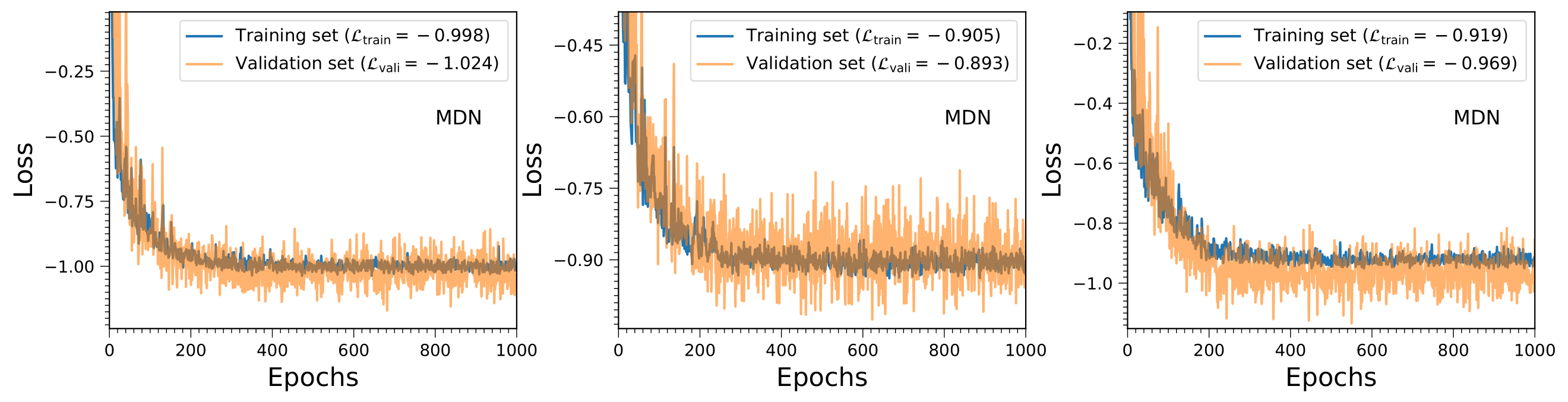}
\caption{Cosmological quantities in the Rastall theory under MDN model. The legends of  (a), (b), and (c) are the same as Fig.~3.}
	\end{figure}
\begin{figure}
	\centering
	(a) \includegraphics[width=0.5\textwidth]{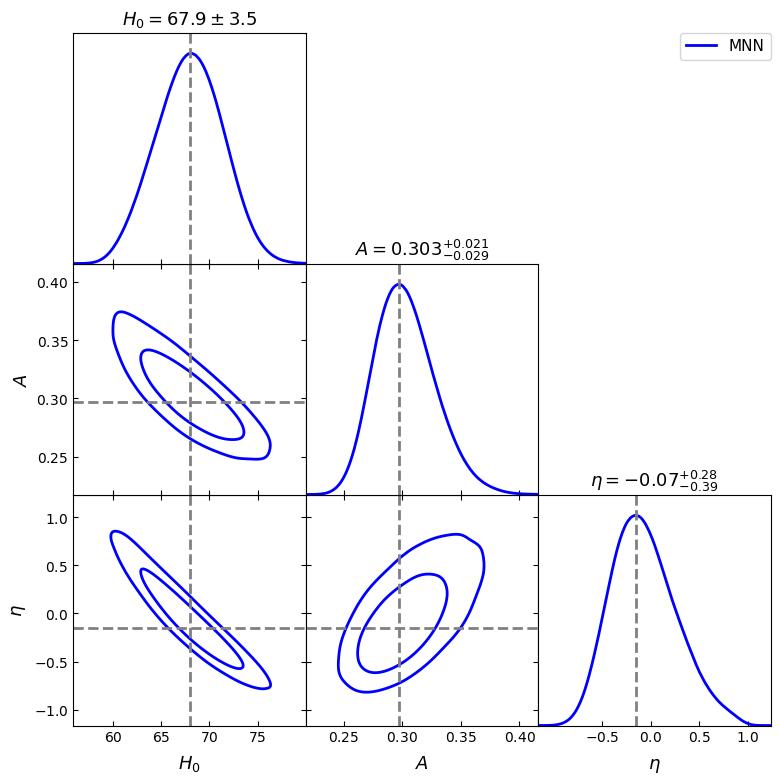}\\
	(b)	\includegraphics[width=0.5\textwidth]{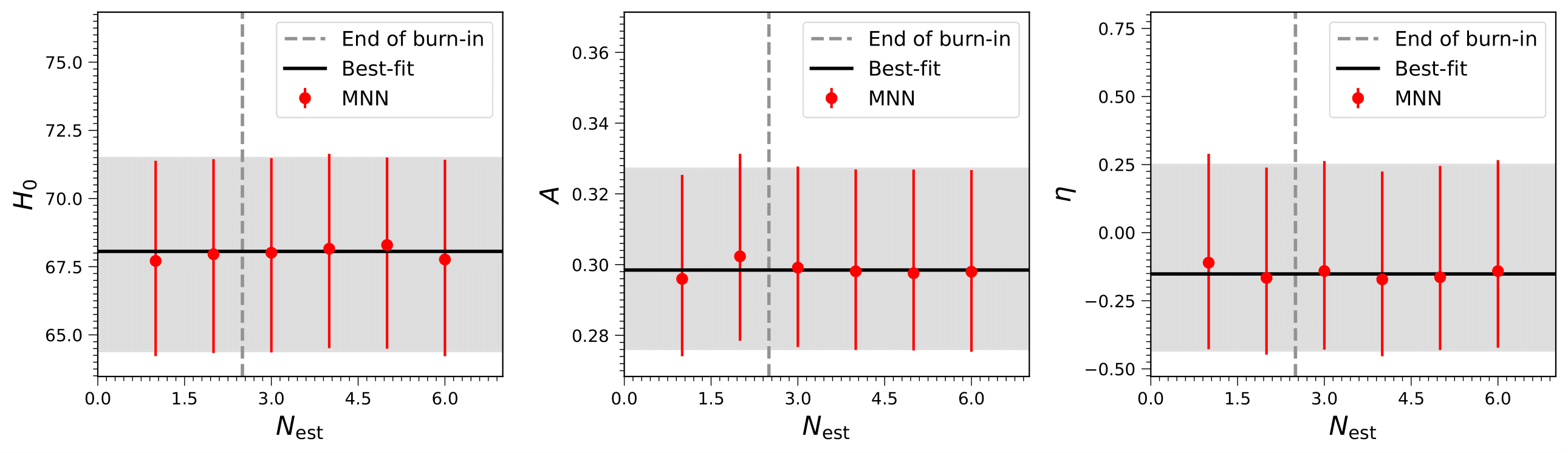}\\
	(c)	\includegraphics[width=0.5\textwidth]{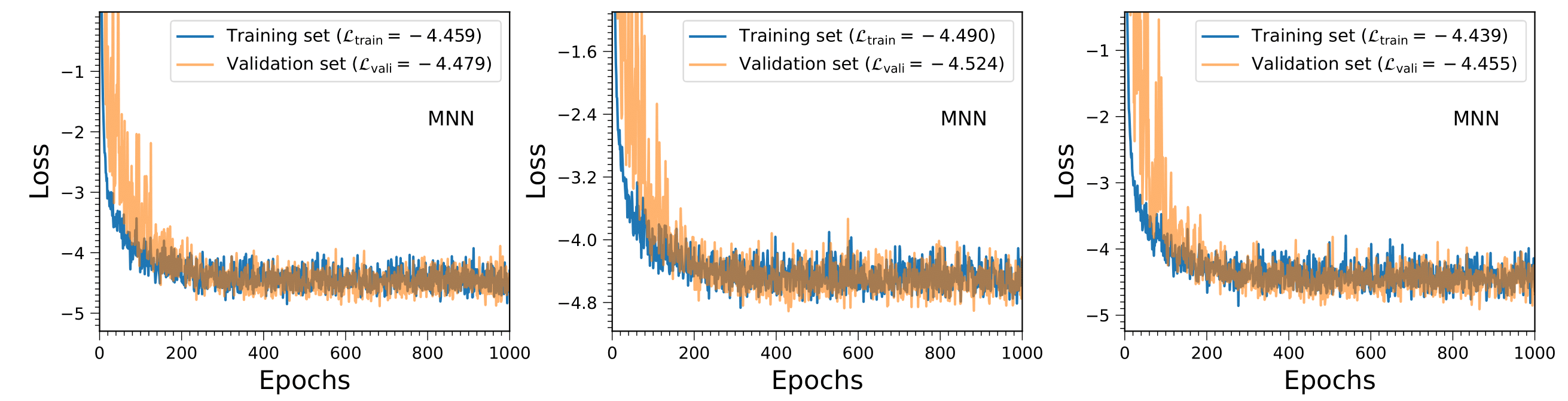}
	\caption{Cosmological quantities in the Rastall theory under MNN model. The legends of  (a), (b), and (c) are the same as Fig.~3.}
\end{figure}

\noindent {\textbf{(i) Artificial Neural Network (ANN):}}

This is, in general, referred to as {\it Neural Network} (NN) which essentially provides a estimating platform to act as a modeling tool. Here, the procedure is as follows: an input layer, may be one or sometimes multiple hidden layers as require due to the demand of the priority or situation, and obviously an output layer. In the case of explicit applications to the cosmological parametric estimation, the generalized methodology is therefore: observational measurements enter through the input layer and are processed through the hidden layers, and thereby eventually producing parametric values of the concerned cosmological model at the expected output layer of the corresponding exit point of the system. Here the procedure involves in each layer operation on a vector containing elements (i.e., neurons) and all these stepwise accept inputs from the previous layer of operation. The computational procedure involves in executing a linear transformation which is coupled with a nonlinear activation function. This normally results with the computed output which is subsequently transmitted to the following layer. Generally, no definitive theoretical guideline exists for selecting the ideal number of neurons to incorporate within each hidden layer. In our investigation, we implement the network design suggested in Ref.~\cite{ref56}, where the neuron count in consecutive hidden layers diminishes progressively as the layer depth increases, facilitating effective feature extraction while minimizing computational demands.

The loss function for ANN method is read as \cite{NIPS2017_addfa9b7}:
\begin{small}
 \begin{equation}
	\mathcal{L} = \mathbb{E} \left( \frac{1}{N} \sum_{i=1}^{N} \left| \theta_i - \hat{\theta}_i \right| \right).
\end{equation}   
\end{small}

The parameter space point \( \theta \) represents the estimated parameters, while the symbol \( N \) denotes the number of cosmological parameters and the target is symbolized by \( \hat{\theta} \) in the training set.\\

\noindent {\textbf{(ii) Mixture Density Network (MDN)}}

This technique properly and effectively integrates an ANN under a mixture modeling framework. This particular mixture modeling technique provides a probabilistic approach on the basic assumption that each data point evolves from a weighted merging for a finite set of probability distributions with unspecified parametric values. Interestingly, these distributions may often belong to different types of probability families which includes Gaussian or beta type distribution. The basic objective of this parametric estimation by using the MDN is to evaluate different unknown parameters which essentially refer to a combined model \cite{ref56}. Hence, an MDN performing Gaussian fraction properly establishes a mapping among the observationally measured datasets and the model parameters which characterizes Gaussian mixture so that the parameters can be optimized through a technique of minimization of the loss function.

The loss function of MDNs methods is read as
\begin{small}
 \begin{equation}
	\mathcal{L} = \mathbb{E} \left[ -\ln \left( \sum_{i=1}^{K} \omega_i \cdot \frac{\exp \left( -\frac{1}{2} (\hat{\theta} - \mu_i)^\top \Sigma_i^{-1} (\hat{\theta} - \mu_i) \right)}{\sqrt{(2\pi)^N |\Sigma_i|}} \right) \right], \label{eqn:loss_mdn}
\end{equation}   
\end{small}
where $\hat{\theta}$ (or $\hat{\boldsymbol{\theta}}$) represents the estimated parameters in both the single- and multi-parameter cases. 

\noindent {\textbf{(iii) Mixture Neural Network (MNN)}}
The Mixture Neural Network (MNN) technique is a posterior distribution that can be exhibited as a merging of various unknown component distributions. The MNN considers this mixture model via an ANN. Here, in this methodology, the steps are as follows: firstly, estimation of the parameters, secondly, by producing samples from the learned mixture reformatting of the posterior distribution is performed. However, in the case of various cosmological parameters whose distributions deviate from the Gaussian feature, essentially multiple mixture components are required to precisely approximate the posterior \cite{ref56}. This obviously increases the computational hazard, as training the network becomes more time-consuming and potentially unstable and hence making the learning of mixture parameters much more challenging.

The modified loss function for model with multiple parameters is obtained as
\begin{small}
\begin{equation}
	\mathcal{L} = \mathbb{E} \left[ - \ln \left( \sum_{i=1}^{K} \omega_i \cdot \frac{\exp \left( -\frac{1}{2} (\boldsymbol{\theta} - \hat{\boldsymbol{\theta}})^\top \boldsymbol{\Sigma}_i^{-1} (\boldsymbol{\theta} - \hat{\boldsymbol{\theta}}) \right)}{(2\pi)^{n/2} |\boldsymbol{\Sigma}_i|^{1/2}} \right) \right]. \label{eqn:loss_mnn}
\end{equation}   
\end{small}

In this setup, the cosmological parameters $\theta$ (or $\boldsymbol{\theta}$) are considered as specific points within the parameter space. The target parameter values, $\hat{\theta}$ (or $\hat{\boldsymbol{\theta}}$), are derived from the training set, while $\boldsymbol{\omega}$ represents the mixture weights.

As an overall observation on the above three ML processes (i.e. ANN, MDN, MNN) it can be expressed as follows: In MCMC process and using the latest combined datasets (including DESI-BAO, CMB, and Union 3.0), the current values of parameters are estimated as $H_{0}=66.945 \pm 1.094$ km s$^{-1}$ Mpc$^{-1}$, $A = 0.304\pm0.008$, $B = 0.696\pm0.008$, and $\eta = -0:322\pm 0:0895$ whereas applying MDN approach the parameters are constrained as $H_{0}=67:040 \pm 0:490$ km s$^{-1}$ Mpc$^{-1}$, $A = 0:302 \pm 0:025$, $B = 0:698 \pm 0:025$, and $\eta = -0:320 \pm 0:210$. These outcomes validate the reliability and effectiveness of the Deep Learning approach used in this paper. In the similar way other techniques like ANN and MNN also constrained the model parameter in strong agreement with the MCMC approach as shown in Tables III and IV.  Thus we can conclude that neural network-based estimation serves as a ``powerful substitute" to the MCMC approach for accurate cosmological modeling within the cosmological framework.

\section{Examination of the model's cosmological features}

In this section, we perform several testings of our model to explore different cosmological attributes with graphical illustrations.

\subsection{EOS parameter}
We have the cosmic density and fluid pressure in the following forms
	\begin{eqnarray}
		\rho &=&  3 \bigg[(1-4 \lambda )H_{0}^2 \left((z+1)^{3}A+B+\eta  \log (z+1)\right)\nonumber\\
		&+& \lambda \left(3 (z+1)^{3} A+\eta \right)H_{0}^2+k (1-2 \lambda ) (z+1)^2\bigg],
	\end{eqnarray}
    
	\begin{eqnarray}
		p &=& -3 (1-4 \lambda )H_{0}^2 \left((z+1)^{3}A+B+\eta  \log (z+1)\right)\nonumber\\
		&+& (1-3 \lambda ) \left(3 (z+1)^{3} A+\eta \right)H_{0}^2-k (1-6 \lambda) (z+1)^2,\nonumber\\
	\end{eqnarray}
and EOS parameter is described as
\begin{eqnarray}
&\omega = \bigg[-3 H_0^2 (1-4 \lambda ) \left[A (z+1)^3+B+\eta  \log (z+1)\right]\nonumber\\
&+H_0^2 (1-3 \lambda ) \left[3 A (z+1)^3+\eta \right]+k (6 \lambda -1) (z+1)^2\bigg]\nonumber\\
&/\bigg[3 \bigg(H_0^2 (1-4 \lambda ) \left[A (z+1)^3+B+\eta  \log (z+1)\right]\nonumber\\
&+H_0^2 \lambda  \left[3 A (z+1)^3+\eta \right]+k (1-2 \lambda ) (z+1)^2\bigg)\bigg].
\end{eqnarray}   

We can note that the Rastall theory based model provides the energy density as positive throughout the evolution of the universe. However, the cosmic pressure initially being positive exhibits a matter-dominated cosmic era. Therefore, for the constraint $\omega > 0$, the EOS parameter features a Big Bang scenario whereas under $\omega = 0$ (representing a non-relativistic matter), a matter-dominated universe can be observed. In this case, the constraint $\omega= 1/3$ (for a relativistic matter) represents a radiation era. Other possibilities are as follows: (i) the model belongs to the quintessence era for the constraint `$-1< \omega \leq 0$', (ii) the EOS parameter provides the $\Lambda$CDM model for $\omega = -1$, and (iii) one can observed a phantom scenario for $\omega<-1$. We have shown the features of $\omega$ in the left panel of Fig. 6 for the presented model under different EOS conditions. Interestingly, this figure exhibits that the model starts from a Big Bang era and then crossing the quintessence phase approaches to the $\Lambda$CDM model universe.

\subsection{Deceleration Parameter (DP)}

The DP can suitably be expressed as 
\[
q = -1 + \frac{(1+z)}{H(z)} \frac{dH(z)}{dz},
\]
so that for the undertaken model, it takes the following form
\begin{equation}
q = \frac{A (1+z)^3 - 2B + \eta - 2\eta \log(1+z)}{2 \left[ A (1+z)^3 + B + \eta \log(1+z) \right]}.
\end{equation}

Under this situation, the model exhibits a clear transition of the universe from a decelerating phase in the past to the current accelerating epoch, signifying the dominance of DE at late times and matter domination in the early universe. For the combined datasets of CC, BAO, and Pantheon, the present value of the deceleration parameter is obtained as $q_0 = -0.595$, with the transition (signature flipping) occurring at a redshift $z_t = 0.685$, as illustrated in Fig. 6 (right panel). The results further indicate that as $z \rightarrow -1$, $q$ asymptotically approaches $-1$, representing a de Sitter-like future. These findings show strong consistency with recent observational analyses \cite{ref76,ref77,Moresco2015,ref89,ref90,ref91,ref92}.

\begin{figure}
	\centering
	\includegraphics[width=0.5\textwidth]{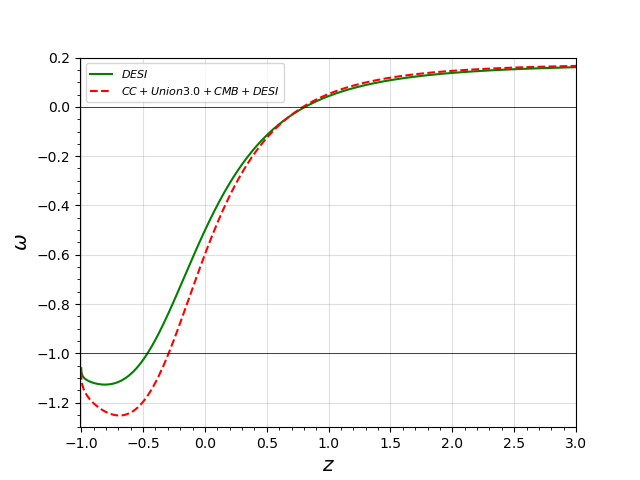}
    \includegraphics[width=0.5\textwidth]{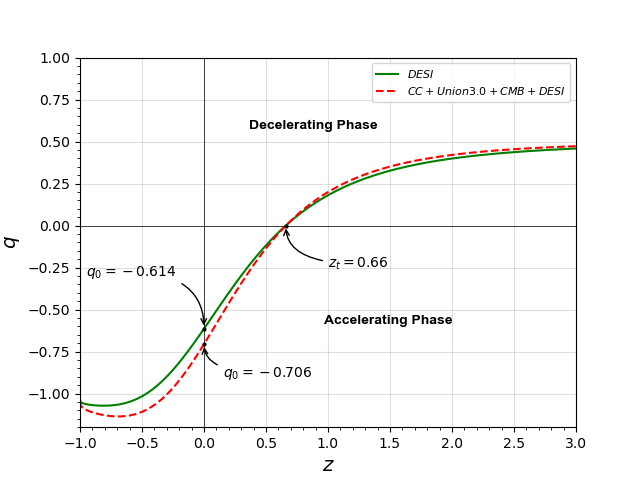}
\caption{$w$ as a function of $z$ (upper panel) and $q$ as a function of $z$ (lower panel). The solid line
(green) shows DESI, and the dotted line (red) depicts CC + Union 3.0 + CMB + DESI datasets.}
\end{figure}

\begin{figure}
	\centering
	(a)\includegraphics[width=0.5\textwidth]{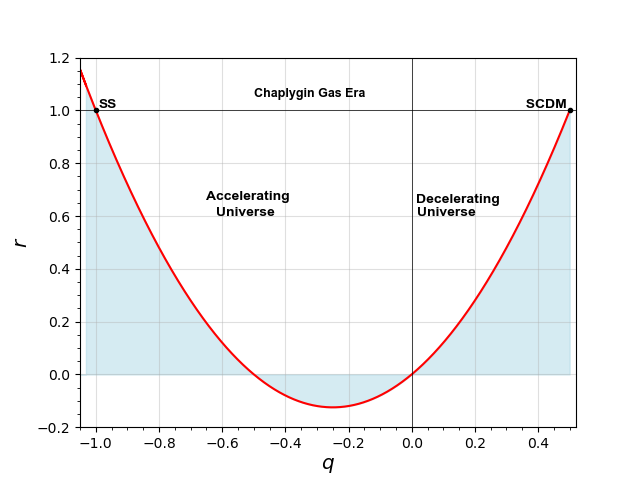}
	(b)\includegraphics[width=0.5\textwidth]{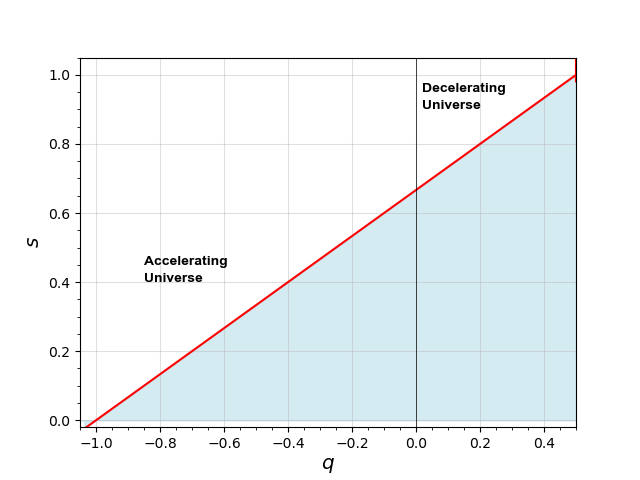}
    (c)\includegraphics[width=0.5\textwidth]{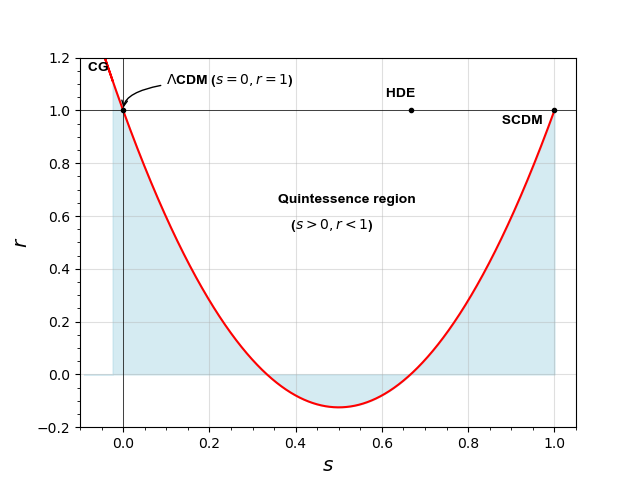}
	\caption{Statefinders diagnostic features: (a) $r$ vs $q$, (b) $s$ vs $q$, and (c) $r$ vs $s$ for joint data set of CC, Union 3.0, CMB, and BAO-DESI datasets.}
\end{figure}
    
\subsection{Statefinders diagnostic}

The statefinder parameters (i.e., $r$ and $s$), can now be introduced for understanding of various DE-based cosmological scenario \cite{ref93,ref94,ref95} via the evolutionary trajectory on the $r-s$ plane, in special connection to the $\Lambda$CDM model \cite{ref61,ref93}. These parameters are essentially a geometrical diagnostic tool which has been developed to distinguish between various DE models. They act even beyond the so-called ordinary {\it acceleration} measured by the Hubble parameter ($H$) and deceleration parameter ($q$). In this scenario $q$ describes about the universe whether it is accelerating, the Statefinder pair ($\{r, s\}$) describes how it is accelerating via the use of the scale factor $a(t)$.

Let us first define $r$ and $s$ as their usual forms, given by
	\begin{eqnarray}
		r&=&2 q^{2}+q - \frac{\dot{q}}{H},
	\end{eqnarray}
	
	\begin{eqnarray}
		s&=&\frac{r-1}{3 (q-\frac{1}{2})},
	\end{eqnarray}
whereas one can find the explicit forms under the Rastall theory theory in Appendix B.

Figure 7(a) represents the evolutionary trajectories for the proposed model in ($r - q$) plane. We can again observe the phase transition of our model by looking into the trajectories of ($r - q$) diagram as ($q$) changes its sign from positive to negative. The evolution of the trajectories for the proposed model initiate with a matter dominated Universe SCDM ($r = 1,~s = 0.5$). As time evolves, the values of $r$ and $q$ start to decline and they attain their minimum position, after which both start to increase towards SS ($r =1,~s = -1$). The progression of the trajectories to SS suggests that our dark energy model may behave like the steady state model in late time.

In the Fig. 7(b), the transition line $q=0$ divide the entire era into two phase. The right phase ($q > 0$) describes the decelerating phase of universe that typically represents the earlier, matter-dominated era of the universe while the left phase ($q < 0$) represents the accelerating phase of the universe which corresponds to the current epoch of our universe, dominated by dark energy.

From Fig. 7(c), it is evident that initially the model taken into consideration exhibits the characteristics like SCDM model ($r = 1, s = 1$). Later on, it  converges to the $\Lambda$CDM scenario ($r = 1, s = 0$) and approaches towards the Chaplygin gas ($r > 1,~s < 0$) era in late passages of the accelerated expanding universe. For the joint sample of CC, CMB, Union 3.0, and DESI datasets, the present values of ($r,~s$) can be estimated to be (0.114, 0.271) which, therefore, displays features resembling the $\Lambda$CDM scenario \cite{ref61,ref94,ref95,ref97}.

\section{Energy Conditions}

The physical viability of the presented model can be verified through the evolution of energy conditions (ECs) \cite{ref98,ref99}. In case of the Rastall theory, the energy conditions can be defined \cite{ref98,ref99,ref100} as follows: (i) Weak energy conditions (WEC), i.e., $\rho \geq 0$, (ii) Dominant energy conditions (DEC), i.e., $\rho - p \geq 0$, (iii) Null energy conditions (NEC), i.e., $\rho + p \geq 0$, and (iv) Strong energy conditions (SEC), i.e., $\rho + 3 p \geq 0$. 

It is to noted that, in general, WEC and DEC are quite satisfiable for all the recognized forms of matters and energies \cite{ref101,ref102,ref103}. However, exotic DE characterized by negative pressure and hence repulsive gravity, is taking the role for the cosmic expansion and SEC violate in the case of DE dominance \cite{ref101,ref102,ref103,ref104}.

Figure 8 demonstrate the feature of various energy conditions in the Rastall theory under a combined dataset of {\color{blue}CC, Union3.0, CMB, and DESI}. In this analysis, DEC is satisfied while NEC and Strong Energy Condition (SEC) violates for the proposed model as depicted in Fig. 8 that validates an accelerated expansion of cosmos in the present era \cite{ref101,ref102,ref103,ref104,ref105}.  

Interestingly, one can note the following two works \cite{Arxiv:2604.16545,Arxiv:2605.03428} where authors have investigated the role of classical energy conditions in static and spherically symmetric spacetimes. 

\begin{figure}
\centering
	(a)\includegraphics[width=0.5\textwidth]{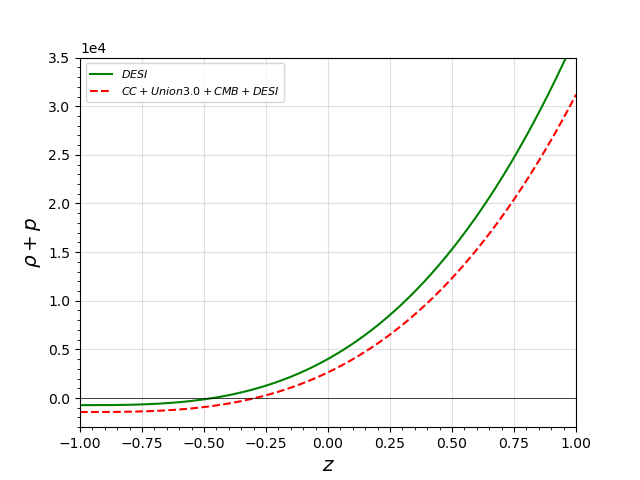}
	(b)\includegraphics[width=0.5\textwidth]{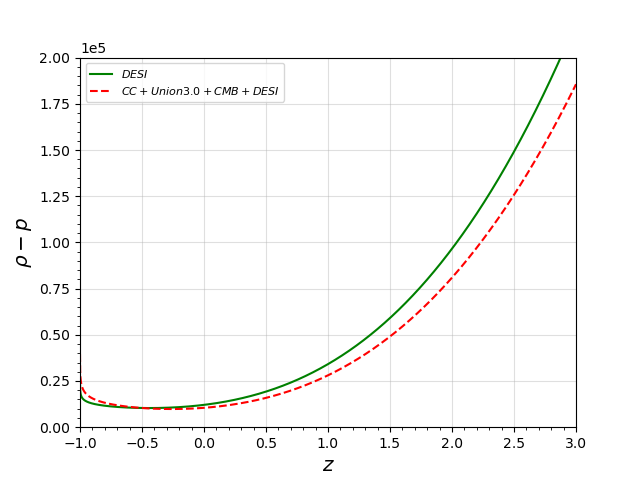}
	(c)\includegraphics[width=0.5\textwidth]{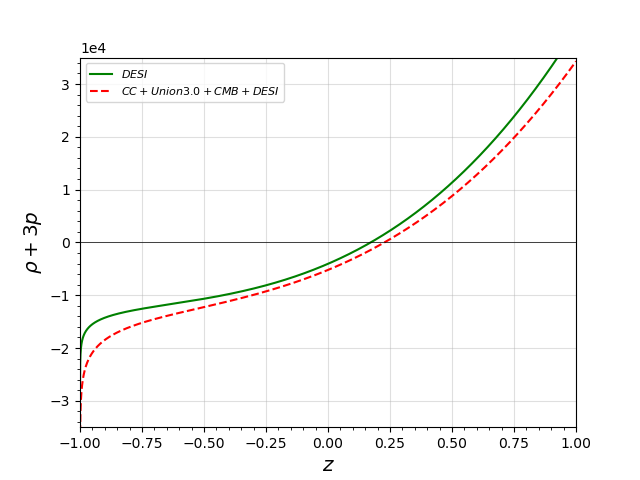}
\caption{Cosmological evolution of energy conditions in the Rastall theory as a function of the redshift $z$ with a combined datasets ({\color{blue}CC, Union3.0, CMB, and DESI}): (a)NEC, (b)DEC, (c)SEC.} \label{fig_energy}
\end{figure}

	\begin{table}[ht]
      \centering
      \small
       \setlength{\tabcolsep}{5pt} 
    \renewcommand{\arraystretch}{1.0} 
	    \caption{Estimated values of the cosmological parameters for the present model by using primary datasets; where machine learning techniques are applied using joint dataset of CC, BAO, and Pantheon.}
	 	\label{tab:t3}
	     \begin{tabular}{ccccc}
    \toprule 
 
                Datasets &  $H_{0}$      &  $A$	 & $B$ &  $\eta$  \\
         &~~~  (km~$\mathrm{s^{-1}}$~$\mathrm{Mpc^{-1}}$) &  	    &     &    \\
				 \hline
				CC &  $68.717 \pm 4.955$ &  $0.3295 \pm 0.0459$  &  $0.6705\pm 0.0459$ &  $-0.1011 \pm 0.7929 $  \\ 
				
				BAO &  $67.434^ {+3.912}_{-4.401}$ &  $ 0.2858 \pm 0.0137$  &  $ 0.7142 \pm 0.0137$ &  $0.3710 \pm 0.2920$  \\ 
				
				Pantheon &  $74.178 \pm 17.093 $ &  $ 0.3614 \pm 0.0292 $  &  $0.6386 \pm 0.0292 $ &  $-0.4631 \pm 0.1786 $  \\ 
				
				CC+BAO+Pantheon &  $69.835 \pm 1.162 $ &  $0.2946 \pm 0.0124$  &  $0.7054 \pm 0.0124$ &  $-0.0744 \pm 0.0881$ \\
				
				ANN &  $68.00\pm 3.40 $ &  $0.302^{+0.021}_{-0.029}$  &  $0.698^{+0.021}_{-0.029}$ &  $-0.070^{+0.290}_{-0.390}$ \\
				
				MDN &  $67.00^{+3.40}_{-3.90} $ &  $0.309^{+0.025}_{-0.031}$  &  $0.691^{+0.025}_{-0.031}$ &  $0.010^{+0.330}_{-0.390}$ \\
				
				MNN &  $67.90\pm 3.50 $ &  $0.303^{+0.021}_{-0.029}$  &  $0.697^{+0.021}_{-0.029}$ &  $-0.070^{+0.280}_{-0.390}$ \\
        \hline
 	    \end{tabular}
	\end{table}    

\begin{table}[ht]
	   \centering
       \small
        \setlength{\tabcolsep}{5pt} 
    \renewcommand{\arraystretch}{1.0} 
	   \caption{Estimated values of the cosmological parameters for the present model by using the latest observational datasets; where machine learning techniques are applied using joint dataset of CC, CMB, DESI, and Union 3.0. }
	 	\label{tab:t4}
	     \begin{tabular}{ccccc}
    \toprule 

	Datasets &  $H_{0}$                                    &  $A$	 & $B$ &  $\eta$  \\
         &~~~  (km~$\mathrm{s^{-1}}$~$\mathrm{Mpc^{-1}}$) &  	    &     &    \\
			\hline
			CC &  $68.717 \pm 4.955$ &  $0.330 \pm 0.046$  &  $0.6705\pm 0.046$ &  $-0.101 \pm 0.793 $  \\ 
			
			DESI &  $72.141 \pm 15.533$ &  $0.3042 \pm 0.0119$  &  $ 0.696 \pm 0.0119$ &  $-0.141 \pm 0.257$  \\ 
			
			Union 3.0 &  $74.047 \pm 1.373$ &  $ 0.312 \pm 0.019 $  &  $0.688 \pm 0.019 $ &  $-0.269 \pm 0.154 $  \\ 
			
			CC+CMB+DESI &  $67.125 \pm 1.430 $ &  $0.302 \pm 0.009$  &  $0.698 \pm 0.009$ &  $-0.321 \pm 0.158$ \\
			
			CC+CMB+DESI+Union 3.0 &  $66.945 \pm 1.094  $ &  $0.304 \pm 0.008$  &  $0.696 \pm 0.008$ &  $-0.322 \pm 0.0895$ \\
			
			ANN &  $67.00\pm 0.470 $ &  $0.301\pm 0.024$  &  $0.699\pm 0.024$ &  $-0.300\pm 0.190$ \\
			
			MDN &  $67.040\pm 0.490 $ &  $0.302\pm0.025$  &  $0.698\pm0.025$ &  $-0.320\pm 0.210$ \\
			
			MNN &  $67.010\pm0.470 $ &  $0.301\pm 0.024$  &  $0.0.699\pm0.024$ &  $-0.300\pm 0.190$ \\
            \hline
			 \end{tabular}
	\end{table}    

In Tables \ref{tab:t3} and \ref{tab:t4}, one can notice from the numerical values of the parameters that overall both the Tables are complying with the observational findings within the valid ranges. However, in a close view of the datasets indicate some minor changes which can be seen in values of model parameters. The latest dataset of DESI-BAO, when clubbed with CC, CMB, and  Union 3.0 determine the Hubble tension $H_0=66.945 \pm 1.094$ which suitably aligned with the findings from the observational evidences reported by Li et al. \cite{li2025desi}.

\section{Information criteria and model selection}

As a final step, we apply the known Akaike Information Criterion (AIC) \cite{Akaike1974stat} and the Bayesian Information Criterion (BIC) \cite{Schwarz1978model}, in order to examine the quality of the fittings and hence the relevant observational compatibility of the scenarios. The AIC is based on information theory, and it is an estimator of the Kullback-Leibler information with the property of asymptotically unbiasedness. Under the standard assumption of Gaussian errors, the corresponding estimator reads as \cite{Anderson2002model,Burnham2004inference}
\begin{equation*}
	{\text{AIC}}  = -2 \ln ({\mathcal{L}_{max}})+2 \alpha +\frac{2 \alpha (\alpha+1)}{N-\alpha-1},
\end{equation*}	
where $\alpha$ denotes the number of free variables in model, $\ln {\mathcal{L}_{max}} $ the maximum likelihood of the datasets and N is the total data points. However, for large number of data points, it reduces to $AIC  -2 \ln ({\mathcal{L}_{max}})+2 \alpha$. On the other hand, the BIC criterion is an estimator of the Bayesian evidence \cite{Anderson2002model,Burnham2004inference}, given by
\begin{equation*}
	{\text{BIC}}  = -2 \ln ({\mathcal{L}_{max}})+2 \alpha \ln N.
\end{equation*}	

To be more precise, a model having $0 \leq \Delta AIC < 2$ and $0 \leq \Delta BIC < 2$ receives strong evidence in favor. In contrast, for $2 < \Delta AIC < 4$ and for $2 \leq \Delta BIC < 6$, the model has average evidence in favor, whereas for $4 < \Delta AIC < 7$ and $6 \leq \Delta BIC < 10$, the model is considered to have less evidence in favor; and, finally, for $ \Delta AIC > 10$ or $ \Delta BIC > 10$, the model receives no significant support since it has no evidence in favor \cite{Liddle2007information}.

Our analysis shows that the $\Lambda$CDM model is preferred by the $\Delta$AIC and $\Delta$BIC criterion. For the combined sample of CC, CMB, Union3.0, and DESI-BAO, we immediately find that $\Delta AIC = 4.592$ and $\Delta BIC = 5.514$. Since, this model shows the differences in the range of $ 4 < \Delta AIC < 7$ and $ 2 \leq \Delta BIC < 6$, therefore, this model has average evidence in favor.

\section{Conclusion}

The present manuscript examines a transitioning cosmological framework implemented through the Rastall theory. We have constructed the relevant field equations within a homogeneous and isotropic spacetime geometry under this theory. To find analytical solutions, we have used a logarithmic representation of the Hubble parameter (see Eq.~\ref{eq10}). Constraints on the model parameters were determined by analyzing observational evidence from different datasets, e.g., BAO, Pantheon, and CC, by using standard MCMC methodology.

To enhance the precision of cosmological parameter estimation and investigation of the current cosmic acceleration phase, we have incorporated an advanced DL technique. By implementing the CoLFI Python package combined with neural architectures including Artificial Neural Networks (ANN), Mixture Density Networks (MDN), and Mixture Neural Networks (MNN), alongside contemporary cosmological observations (CMB, BAO, DESI, and Union 3.0), we achieved efficient parameter determination. This methodology significantly has enhanced our comprehension of the provisional likelihood functions obtained from observational evidence and their associated posterior probability distributions (illustrated in Figs.~3--5 as well as in Figs. 9--11). 

Under this motivation we have performed statefinder diagnostic evaluation which reveals that our model demonstrates behavior analogous to the $\Lambda$CDM paradigm during the current epoch, while exhibiting certain deviations from complete conformity (illustrated in Fig.~7). We note that there is a violation of SEC in the Rastall theory framework which therefore provides evidence in support of the accelerated expansion phase of the universe in contemporary times (illustrated in Fig.~8). Therefore, our examination of fundamental cosmic quantities -- which include the pressure, energy density, EOS  parameter, and DP -- reveals that the presented model produces physically meaningful and consistent outcomes (illustrated in Figs.~6--8).

Furthermore, employment of the hyper-ellipsoid constraints contributed to enhance stability during neural network optimization as well as improved the accuracy of the concerned parameter determination. A comparative evaluation among the modern NN-based approaches and that of the traditional MCMC methodology revealed that the MNN technique produced results in close agreement with MCMC outcomes, thus confirming its effectiveness and dependability for cosmological parameter inference. {\color{blue}Our analysis shows that MNN performs best among the evaluated ML techniques and in comparison with the MCMC approach. The MNN technique produces results in strongest agreement with those obtained from the traditional MCMC method, validating its reliability and effectiveness over the other approaches. For example, when using the latest combined dataset (CC + CMB + DESI + Union 3.0), the parameter estimates between MCMC and MNN are nearly identical: 

MCMC: $H_0 = 66.945 \pm 1.094 \text{ km s}^{-1}\text{Mpc}^{-1}$, $A = 0.304 \pm 0.008$, $\eta = -0.322 \pm 0.0895,$  

MNN: $H_0 = 67.010 \pm 0.470 \text{ km s}^{-1}\text{Mpc}^{-1}$, $A = 0.301 \pm 0.024$, $\eta = -0.300 \pm 0.190.$
}

The presented cosmological model successfully accounts for the late-time cosmic acceleration without invoking any mysterious DE, thereby circumventing complications related to the erstwhile cosmological constant. This emphasizes the significance of incorporating current observational constraints when modeling within the Rastall theory paradigm. Our results demonstrate that neural network approaches provide an effective methodology for determining cosmological parameters, offering a practical complement to MCMC methodologies. The Rastall theory paradigm enables investigations of diverse gravitational astrophysics and their influence on cosmic expansion dynamics utilizing ANN, MNN, and MDN frameworks. Evidently, this investigation forms part of an extensive program merging cosmology with ML, showcasing ML-based technical capability for addressing complex cosmic dynamics and expansion phenomena  (illustrated in Figs.~3--5). 

\begin{table}[ht]
	   \centering
       \small
       \setlength{\tabcolsep}{5pt} 
    \renewcommand{\arraystretch}{1.0} 
	   \caption{Comparison of the current value of the Hubble parameter:  the present model vs other models in the literature.}
	 	\label{tab:t5}
	     \begin{tabular}{ccccc}
    \hline \hline 

	Datasets &  $H_{0}$ (km~$\mathrm{s^{-1}}$~$\mathrm{Mpc^{-1}}$)   \\
			\hline
			DESI &  $72.141 \pm 15.533$~\footnote{{\it Data derived from the present model}} &  $73.43 \pm 2.97$~\cite{LH2025JCAP}  &  $68.4^{+1.0}_{-0.8}$~\cite{Guo2024arxiv} & $68.53 \pm 0.80$~\cite{Adame2025desi}   \\ 
			
			Union 3.0 &  $74.047 \pm 1.373$~\footnote{{\it Data derived from the present model}} &  $73.38^{+3.12}_{-2.82}$~\cite{LH2025JCAP}  &  $74.912^{+2.015}_{-1.805}$~\cite{DK2025Arxiv} &  $67.93 \pm 0.42$~\cite{MAS2025Arxiv}  \\ 
			  \hline
			 \end{tabular}
	\end{table}    

In Table \ref{tab:t2}, we have shown the Hubble parametric values based on CC and BAO methodology. Tables \ref{tab:t3} (derived values of the parameters for the presented model by using primary datasets) and \ref{tab:t4} (derived values of the parameters for the presented model by latest observational datasets) present the quantitative values of model parameters. While both tables demonstrate reasonable agreement with observations, however derived from the most recent observational data exhibits superior consistency with observational signatures overall. We have done a bit comparative study in Table \ref{tab:t5} by showing the numerical data of $H_0$: our case of the presented model vs a few available data from the literature. A close observation on this Table \ref{tab:t5} shows an overall confirmity and validity of our model in comparison to others investigations.  

Essentially, from our investigations by using ML techniques (Tables \ref{tab:t3} and \ref{tab:t4}), we observe that the Hubble tension for the present model is nearly identical to the estimated value by MCMC for the combined dataset (i.e., CMB, CC, Union 3.0, and DESI-BAO). So, we note that ML approaches (i.e. ANN, MDN, and MNN) work in a better way to get statistical measure of the model parameters. 

Finally, we can put forward the following two aspects:\\
(i) {\it Data analytic point of view}: The current study has analyzed the Hubble parameter together with the latest DESI-BAO data and DE Survey observations through statistical analysis (particularly the MCMC method) as a primary platform and ML methodology as higher level application. 

In our future research program, we intend to analyze gravitational wave data utilizing ML approaches, which have demonstrated superior performance relative to traditional MCMC methods and may therefore reveal novel physical insights and open new research directions within our adopted framework. In this direction, a different type of application is already available \cite{Bhardwaj2025}.

(ii) {\it Theoretical point of view}: A few interesting works are available in the literature~\cite{Silva2013,Sharif2025} where authors have shown that Rastall gravity may produce Bouncing cosmological universe. This possible aspect has not dealt with in the present investigation and can be thought for our future projects. Even there are possibilities to apply the Rastall gravity along with the Rainbow gravity in a more versatile approach \cite{Battista2024}. 
\\

\section*{Acknowledgments}
SR sincerely acknowledges the support of the Inter-University Centre for Astronomy and Astrophysics (IUCAA), Pune, India, through its Visiting Research Associateship Programme, during which part of this study was carried out. 

\section*{Acknowledgments}
The authors extend their appreciation to the Deanship of Research and Graduate Studies at King Khalid University for funding this work through the Large Research Groups Program under grant number RGP2/12/47. SR sincerely acknowledges the support of the Inter-University Centre for Astronomy and Astrophysics (IUCAA), Pune, India, through its Visiting Research Associateship Programme, during which part of this study was carried out. The research work of KB was partially supported by the JSPS KAKENHI Grant Numbers 24KF0100, 25KF0176.

\makeatletter
\renewcommand{\@biblabel}[1]{[#1]}
\makeatother

\end{document}